\documentclass[english,12pt,a4paper]{article}
\usepackage{authblk}
\usepackage[text={17.0cm,23.7cm}]{geometry}
\usepackage[T1]{fontenc}
\usepackage[utf8]{inputenc}
\usepackage{babel}
\usepackage{graphicx}
\usepackage{tabularx}
\usepackage{amsmath}
\usepackage{multirow}
\usepackage{booktabs}
\usepackage{amssymb}
\usepackage[dvipsnames]{xcolor}
\usepackage{yfonts}
\usepackage{ulem}
\usepackage{enumitem}
\usepackage{mdframed}
\usepackage{xcolor}
\usepackage[style=numeric-comp, sorting=none]{biblatex}
\usepackage{float}
\usepackage{mathrsfs}

\newcommand{\mdm}{m_{\text{DM}}}
\newcommand{\dER}{\text{d}E_R}
\usepackage{biblatex}
\usepackage[colorlinks=true, allcolors=blue]{hyperref}

\begin{document}

	\title{\vspace{-2cm}
		\vspace{0.6cm}
		\textbf{Information divergences to parametrize astrophysical uncertainties in dark matter direct detection}\\[8mm]}
	\author[1,2,3]{Gonzalo Herrera}
	\author[2]{Andreas Rappelt}
 
	\affil[1]{\normalsize\textit{Center for Neutrino Physics, Department of Physics, Virginia Tech, Blacksburg, VA 24061, USA}}
	\affil[2]{\normalsize\textit{Physik-Department, Technische Universit\"at M\"unchen, James-Franck-Stra\ss{}e, 85748 Garching, Germany}}
	\affil[3]{\normalsize\textit{Max-Planck-Institut f\"ur Physik (Werner-Heisenberg-Institut), F\"ohringer Ring 6,80805 M\"unchen, Germany}}

	\date{}
	\maketitle
	
	\begin{abstract}
     Astrophysical uncertainties in dark matter direct detection experiments are typically addressed by parametrizing the velocity distribution in terms of a few uncertain parameters that vary around some central values. Here we propose a method to optimize over all velocity distributions lying within a given distance measure from a central distribution. We discretize the dark matter velocity distribution as a superposition of streams, and use a variety of information divergences to parametrize its uncertainties. With this, we bracket the limits on the dark matter-nucleon and dark matter-electron scattering cross sections, when the true dark matter velocity distribution deviates from the commonly assumed Maxwell-Boltzmann form. The methodology pursued is general and could be applied to other physics scenarios where a given physical observable depends on a function that is uncertain.
	\end{abstract}
		
	\section{Introduction}

In the theoretical description of many physical processes, an observable $F$ is a functional of an input function $f$, $F[f]$. For example, the dark matter scattering rate in a direct detection experiment is a functional of the local dark matter velocity distribution; or the gamma-ray/neutrino flux at Earth from dark matter annihilation in the Universe is a functional of the dark matter density distribution. The function $f$ may be calculated from first principles, or be determined from experiments. In either case, it is subject to theoretical and experimental uncertainties, that impact the theoretical prediction of the observable $F[f]$. 

\begin{figure}[H]
	  \centering
       \includegraphics[width=0.7\linewidth]{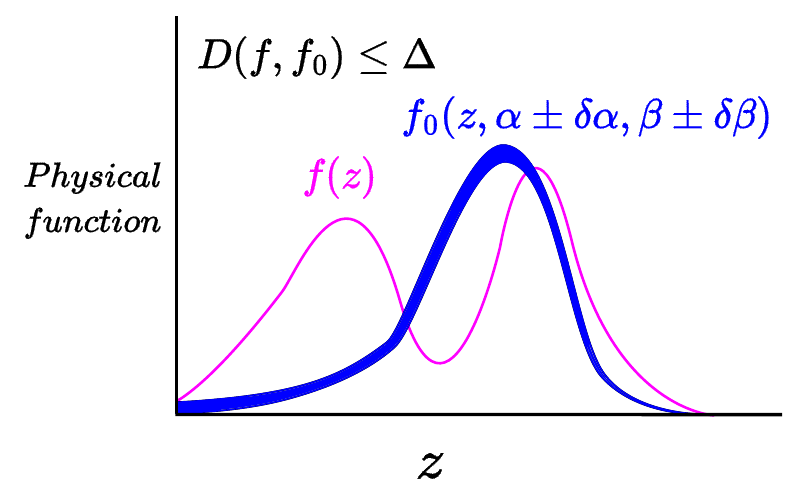}
		\caption{The blue lines represent the family of functions obtained by varying the function parameters $\alpha$ and $\beta$ around a central value. The magenta line represents a function that deviates at most a certain value $\Delta$ from the central value of the function $f_0$, given a distance measure $D$ parametrizing the deviation among the two functions. It can be clearly seen that varying the parameters $\alpha, \beta$ of the central function $f_0$ wouldn't capture the parametric form of the function $f$.}
		\label{fig:diagram}
	   \end{figure}

In order to bracket the uncertainty in the theoretical prediction of $F[f]$, it is common in the literature to parametrize $f$ in terms of a number of parameters, that vary around some central values. We propose an alternative which consists in using methods of convex optimization to optimize over all functions lying within a given distance measure from a central function, see Fig \ref{fig:diagram}.

As distance measures to parametrize the deviation from the distribution, we use information divergences, widely used in mathematics and statistics to quantify differences between probability distributions \cite{kullback1951, Rnyi1961OnMO, Csiszar1964}. Information divergences are a type of distance measures that quantify the difference between two probability distributions, often relying on the information entropy of a distribution $f(z)$ over a continuous random variable $z$, defined as

\begin{equation}
H[f] = \int dz f(z) \log (f(z))
\end{equation}
where the base of the logarithm determines the units, \textit{e.g} base 2 corresponds to bits, and base $e$ corresponds to nats.

In particular, the method that we propose consists in optimizing the functional $F$ under some given constraints on $f$, among which one sets a maximal distance $\Delta$ between the central function $f_0$ and the optimal function $f$, $D(f,f_0) \leq \Delta$, given by information divergences. The problems of this class are not always convex neither simple to solve while guaranteeing that the found optimal solution is global. Nonetheless, many of these problems fall in some of the categories of convex problems such as linear programming (LP), quadratic programming (QP), second order cone programming (SOCP), semi-definite programming (SDP), or exponential programming (EXP) \cite{boyd_vandenberghe_2004}. Such problems can be solved with convex optimization techniques that work on problems with objective functions that are convex, and therefore have a feasible region allowed by the problem constraints. In this context, there are multiple softwares able to solve this problems using interior-point methods. Interior-point methods work by sampling the interior of the feasible region, unlike previous algorithms (like the simplex) which transveres the boundary of the feasible region. A very compelling example of these softwares is \texttt{CVXPY} \cite{diamond2016cvxpy}, since it embeds different codes with a narrower applicability to provide a software able to identify and solve a large number of convex optimization problems, via disciplined convex programming \cite{Grant06disciplinedconvex}.
	 
We develop a methodology that allows to bracket the impact of theoretical uncertainties in a given experiment by making use of information divergences and convex optimization techniques, and apply our method to the field of direct dark matter searches. In direct dark matter searches, the rate of dark matter induced scatterings in a detector is a functional of the dark matter velocity distribution. The form of the dark matter velocity distribution, which is as of today unknown, plays a crucial role when interpreting direct dark matter searches. In order to interpret the outcome of direct dark matter searches, one usually assumes the Standard Halo Model. On the other hand, several works showed that deviations from the SHM are expected to be sizable, having an impact in nuclear recoil searches, \textit{E.g} \cite{Vogelsberger:2008qb, Baushev:2012dm, Schaye_2014, Green:2017odb, Necib:2018iwb, evans2018shm, O_Hare_2018, Mandal:2018efq, Besla:2019xbx,Buch:2019aiw,Herrera:2021puj, Herrera:2023fpq, Smith-Orlik:2023kyl}, and in electron recoil searches, \textit{E.g} \cite{Hryczuk:2020trm, Buch:2020xyt, Maity:2020wic,Radick:2020qip, Maity:2022enp, Herrera:2021puj, Herrera:2023fpq}. This source of uncertainty has been treated with analysis frameworks to interpret direct dark matter searches with halo-independent methods and more sophisticated integration techniques \cite{Fox:2010bz, Fox:2010bu, Gondolo:2012rs, Herrero-Garcia:2012arz, DelNobile:2013cta, Frandsen:2013cna, Bozorgnia:2014gsa, Feldstein:2014gza,Blennow:2015oea, Anderson:2015xaa, Gelmini:2016pei, Kahlhoefer:2016eds, Ibarra:2018yxq, Ibarra:2017mzt, Catena:2018ywo, fowlie, Chen:2021qao, Chen:2022xzi, Kang:2022zqv, Kang:2023gef,Bernreuther:2023aqe, Lillard:2023cyy, Lillard:2023nyj}.

The paper is organized as follows. In section \ref{sec:FuncOpt1} we describe the mathematical form of the optimization problem that we aim to solve. In section \ref{sec:Distances} we introduce a zoo of distance measures used to parametrize the uncertainties on the dark matter velocity distribution. In section \ref{sec:Distances_VDF}, we quantify the deviations from the SHM for different distance measures and models of the dark matter velocity distribution present in the literature. In section \ref{sec:Nuclear}, we derive upper limits for nuclear recoil searches for different choices of the distance measure. In section \ref{sec:Electron} we derive upper limits for electron recoil searches for different choices of the distance measure, and study the projected complementarity of XENON1T and SENSEI in the hypothetical scenario that one experiment claims a detection while the other does not. In particular, we quantify the compatibility of a positive signal in one experiment and a null signal in the other when the dark matter velocity distribution deviates from the Maxwell-Boltzmann a maximum value.
  
\section{Functional optimization}\label{sec:FuncOpt1}
	
	We consider the objective functional $\mathcal{F}[f]$ with $f$ a real function with domain $S$. In some physical applications, one aims to find maximum and minimum value of the objective functional $\mathcal{F}$, subject to a number of constraints that are also functionals of $f$. Specifically, the optimization problem can be formulated as: 
    \begin{align}
		\label{eq:FuncOpt1}
		\text{optimize}~~~~~&\mathcal{F}[f]\nonumber \\
		\text{subject to}~~~~~&\mathcal{C}_\alpha[f]= 0,~~~\alpha=1,\dots,p\nonumber\\
		&\mathcal{C}_\alpha[f] \leq 0,~~~\alpha=p+1,\dots,p+q
	\end{align}
 
	where in full generality the constraints consist of $p$ equalities and $q$ inequalities\footnote{Constraints of the form ${\cal C}_\beta\geq 0$ amount the redefinition ${\cal C}_\beta\rightarrow -{\cal C}_\beta$.}. The domain of the function $f$ can be discretized in $N$ points, $z_i$, $i=1,\cdots, N$, so that the functionals ${\cal F}[f]$ and ${\cal C}_\alpha[f]$ become functions of $N$ variables $f_i\equiv f(z_i)$. The optimization problem over the continuous function $f(z)$ then turns into an optimization over a (large) number of discrete variables,
	\begin{align}
		\label{eq:FuncOptDis1}
		\text{minimize}~~~~~&F\big(f_1,\cdots, f_N\big)\nonumber \\
		\text{subject to}~~~~~&C_\alpha\big(f_1,\cdots, f_N\big)= 0,~~~\alpha=1,\dots,p\nonumber\\
		&C_\alpha\big(f_1,\cdots, f_N\big)\leq 0,~~~\alpha=p+1,\dots,p+q
	\end{align}
 
 We are able to state the described problem in a convex form for several objective functions and constraints on the function $f$. With this, the objective function as well as the constraints solely depend on the vector $f_{i}$. One can therefore formalize a convex optimization problem in terms of matrix inequalities, either linear, quadratic (See \cite{Rappelt:2020tch}), or even logarithmic and exponential (See Appendix \ref{sec:A2}), which can then be solved with some numerical softwares, such as \texttt{CVXPY}.

\section{Distance measures: Information divergences}\label{sec:Distances}

The form of the functionals ${\cal F}[f]$ in the objective function of Eq. \ref{eq:FuncOpt1} involve the integral of a function $f$ over a domain, and do not allow to constrain the local properties of the function $f$. The function that solves the optimization problem of Eq \ref{eq:FuncOpt1} could therefore be misleading from the physical point of view, for example if the solution is discontinuous, or presents sharp features such as delta functions (see \textit{E.g} \cite{Ibarra:2018yxq, Rappelt:2020tch}). In this case, the bracketing of the functional ${\cal F}[f]$, although strict from the mathematical point of view, could be too conservative from the physical point of view, since it is unlikely that the upper (lower) limit on ${\cal F}[f]$ will be saturated in reality. 

A more realistic bracketing of the functional  ${\cal F}[f]$ would require to include in the optimization problem  additional restrictions on the function $f$ (denoted by $\mathcal{C_{\alpha}}$ in Eq. \ref{eq:FuncOpt1}), motivated by theoretical considerations or by experimental results. For instance, one could restrict the family of functions $f(z)$ entering in the optimization to be close from a theoretically motivated function $f_0(z)$ and/or to lie within a given range $f_{\rm min}(z)\leq f(z) \leq f_{\rm max}(z)$. A special case would be
\begin{align}
f(z)\geq 0,~~{\rm for~all}\, z
\end{align}
as required if $f(z)$ is probability distribution.

Furthermore, we aim to find an appropriate distance measure to parameterize the deviation of the true dark matter velocity distribution from the Maxwell-Boltzmann (see Table \ref{tab:distances} for a compilation of distance measures considered in this work). A common parametrization of the distance between functions is given by the $L^p$-norm, defined as
	\begin{equation}
		\|f\|_p=\left(\int\mathrm{d}z\,|f(z)|^p\right)^{1/p}.
	\end{equation}
    such as the common $L^{1},L^{2}$ and $L^{\infty}$ norms. In the following, we'll discuss that an approppiate alternative to $L^{p}$-norms are information divergences. In information theory, divergence refers to a weaker definition of distance between two probability distributions on a given domain. Although not a metric, information divergences \cite{kullback1951} offer an interesting alternative to quantify the distance of two distributions. Information divergences are non-symmetric between two distributions $f$ and $f_0$. We assume that $f$ is the true distribution and $f_0$ is the reference distribution. 

An important class of divergences are the $f$-divergences \cite{Rnyi1961OnMO,Csiszar1964}, which are in general defined as
	\begin{align}
		D_{f}[f,f_0] = \int dz \, f_0(z)\,F\left(\frac{f(z)}{f_0(z)}\right)-F(1)
	\end{align}
where $F$ is a convex function with minimal value $F(1)$. This definition ensures the distance between two identical functions to be 0. It should be noted that we are abusing of notation here. In reality, the functions $f(z)$ may correspond to probability densities instead of probability distributions, with inverse units to $z$. The $f$-divergences are, however, dimensionless, regardless of $f$ being probability distributions or probability densities. Some examples of $f$-divergences used in this work are the Vajda divergences, the \textit{$\chi^{2}$}-divergence and the KL-divergence. The Vajda divergences are a subclass of $f$-divergences where $F(z)=|z-1|^\alpha$ with $\alpha\geq 1$.
As $F(1)=0$, the so called $\chi^\alpha$-Vajda divergence reads
	\begin{align}
		D_{\chi^\alpha}[f,f_0] = \int dz \,|f(z)-f_0(z)|^\alpha\,f_0(z)^{1-\alpha}\,.
	\end{align}
There are two interesting special cases which can be converted into generalized inequalities.
First, we recover the $L^1$ norm when choosing $\alpha=1$
	\begin{align}
	D_{\chi^1}[f,f_0] = \int dz \,|f(z)-f_0(z)|\,,
	\end{align}
Informationally, this is the largest possible difference between the probabilities that the two probability distributions can assign to the same event. Another interesting example is $\alpha=2$ for which the $\chi^\alpha$-Vajda divergence is equal to the Pearson $\chi^2$ measure, which corresponds to the specific choice of $F(z)=z^2-1$ and reads
	\begin{align}
		D_{\chi^2}[f,f_0] = \int dz \, \frac{|f(z)-f_0(z)|^2}{f_0(z)}\,.
	\end{align}
The reverse case will be a convex functional and it is called the Neyman $\chi^{2}$-divergence, obtained by taking $F(z)=\frac{1}{z}-1$ and reads
\begin{equation}\label{eq:chisquared}
    \centering 	D_{\chi^2}[f_0,f] = \int dz \, \frac{|f(z)-f_0(z)|^2}{f(z)}\,.
\end{equation}

The Kullback-Leibler (KL) divergence is obtained by taking $F(z)=z\,\mathrm{log}z$ and is then defined as \cite{kullback1951}\footnote{Note that the integrand of the KL divergence can take negative values due to the presence of the logarithm of the ratio between distributions, however, the KL-divergence obtained after integrating over the whole domain is positive-defined. This can be probed by Gibbs inequality $-\int dz \,  f_0(z) \, \mathrm{log}(f(z)) \leq-\int dz \, f(z)\, \mathrm{log} (f_0(z))$.}.

\begin{equation}\label{eq:KLdivergence}
     D_{KL}[f,f_0]= \int dz \, f(z)\textit{ } \mathrm{log} \left (\frac{f(z)}{f_0(z)}  \right )\,.
\end{equation}

The KL-divergence is a well motivated distance measure to compute deviations with respect to Gaussian or Maxwellian distributions, which are common in physics. This is due to the fact that the KL-divergence can be expressed as
    \begin{equation}
        D_{KL}[f,f_0]= -\int dz \, f(z)\textit{ } \mathrm{log}(f_0(z)) + \int dz \, f(z)\textit{ } \mathrm{log}(f(z)) = H[f,f_0]-H[f]
    \end{equation}
where $H[f,f_0]$ is the cross-entropy of $f$ and $f_0$ and $H[f]$ is the entropy of $f$. Therefore, the KL-divergence is interpreted as the relative entropy of $f$ with respect to $f_0$. From a bayesian inference point of view, $D_{KL}[f,f_0]$ is a measure of the information gained when revising one's beliefs from the prior probability distribution $f_0$ to the posterior probability distribution $f$. This interpretation motivates its use as a constraint in our functional optimization problems, where we want to parameterize the deviation with respect to a given prior distribution.

    		\begin{table}[H]
		\begin{center}
			\begin{tabular}{c|c}
				\toprule
				\toprule
				Distance measures & $ D[f,f_{0}]$ \\
				\bottomrule
				\toprule
				$L^{1}$-norm & $\displaystyle{\int\text{d}z \left |f(z)-f_{0}(z)  \right |}$  \\
				\vspace{2mm}
				$L^{\infty}$-norm & $ \mathrm{max} \big( \left |f(z)-f_{0}(z) \right |\big )$  \\
				\vspace{2mm}
				$\chi^{2}$-divergence & $\displaystyle{\int\text{d}z \frac{\big ( f_{0}(z)-f(z) \big )^{2}}{f_{0}(z)}}$  \\
				\vspace{2mm}
                KL-divergence & $ \displaystyle{\int\text{d}z  f(z)\log \frac{f(z)}{f_{0}(z)}}$  \\
				\bottomrule
				\bottomrule
			\end{tabular}
		\end{center}
		\caption{Distance measures considered in this work, where $f_{0}$ is a reference (prior) distribution, from which a maximal distance to the true distribution $f$ is set.}
		\label{tab:distances}
	\end{table}

Divergences can be related to each other. For our purposes, the relations help to understand the scaling between them and to constrain the true set of feasible dark matter velocity distributions that would not result in a dark matter signal in a certain experiment and deviate a certain amount from the Maxwell-Boltzmann. For example, the KL-divergence presents a lower bound in terms of the $L^{1}$ distance, which is as well lower bounded by the Hellinger distance $H[f,f_0]$

\begin{align}\label{eq:ineq1}
  2H^{4}[f,f_0] \leq \frac{1}{2}\left \|f-f_0  \right \|_{1}^{2} \leq D_{KL}[f,f_0].
\end{align}
where the Hellinger distance is defined as
\begin{align}
    \frac{1}{2}\int\text{d}z \Big[\sqrt{f(z)}-\sqrt{f_{0}(z)}\Big]^{2}.
\end{align}

Such relations among distance measures allow to understand their relative behavior as parametrization of the uncertainties present in a function encoded in a physical observable.

\section{Distance between velocity distributions}\label{sec:Distances_VDF}
With the formalism developed in the previous section, it is possible to determine the deviation of the true dark matter velocity distribution to a given reference velocity distribution.
Concretely, we calculate this for the velocity distributions extracted from the Aquarius and Eagle simulations \cite{Vogelsberger:2008qb, Bozorgnia:2016ogo}, determined using Eddington's formula\footnote{The phase-space distribution of the dark matter particles can be obtained from Eddington's formula as $\mathcal{F}(\mathcal{E})=\frac{1}{\sqrt{8} \pi^2}\left[\int_0^{\mathcal{E}} \frac{d \Psi}{\sqrt{\mathcal{E}-\Psi}} \frac{d^2 \rho}{d \Psi^2}+\frac{1}{\sqrt{\mathcal{E}}}\left(\frac{d \rho}{d \Psi}\right)_{\Psi=0}\right]$, where $\Phi$ is the gravitational potential of the system, $\Psi(r) \equiv-\Phi(r)+\Phi(r=\infty), \mathcal{E} \equiv-E+\Phi(r=\infty)=$ $\Psi(r)-\frac{1}{2} v^2$, and $\rho(r)$ is the density profile. The velocity distribution is then obtained with this method at a given radius $r$ as $f(v)=\frac{\mathcal{F}}{\rho(r)}$.} \cite{Mandal:2018efq} and reconstructed from the motion of stars \cite{Necib:2018iwb, McCabe:2013kea, O_Hare_2018}.
We confront each of those distributions to the Standard Halo Model, which models the velocity distribution as a Maxwell-Boltzmann distribution
        \begin{align}
    		f_\text{SHM}(\vec{v}) = \frac{1}{N_\text{esc}\,(2\pi\,\sigma_v^2)^{3/2}}\,\exp\left(-\frac{\vec{v}^2}{2\,\sigma_v^2}\right)\,,
    	\end{align}
where $\sigma_v\approx 156\,\text{km/s}$ is the velocity dispersion \cite{Kerr:1986hz, Green:2011bv}.
The normalization constant $N_\text{esc}$ depends on the escape velocity $v_\text{esc}\approx 544\,\text{km/s}$ \cite{Smith:2006ym, Piffl:2013mla} and is given by
    	\begin{align}
    		N_\text{esc} = \text{erf}\left(\frac{v_\text{esc}}{\sqrt{2}\sigma_v}\right)-\sqrt{\frac{2}{\pi}}\,\frac{v_\text{esc}}{\sigma_v}\,\exp\left(-\frac{v_\text{esc}^2}{2\sigma_v^2}\right)\,.
    	\end{align}
In Tab. \ref{tab:CompNorms}, we give the values of the distance of some velocity distributions to the SHM.
In each case, we consider the angular averaged distributions in the solar rest frame
    	\begin{align}
    		f_{0}(v) = v^2\,\int\text{d}\Omega_v\,f(\vec{v}+\vec{v}_\odot)\,,
    	\end{align}
where $\vec{v}_\odot$ is the velocity of the Sun with respect to the Galactic rest fame.
The velocity of the Sun is composed of the motion of the local standard of rest (LSR) and the peculiar motion of the Sun with respect to the LSR, i.e.
    	\begin{align}
    		\vec{v}_\odot = \vec{v}_\text{LSR} + \vec{v}_{\odot, \text{pec}}\,.
    	\end{align}
Concretely, the motion of the LSR is given by $\vec{v}_\text{LSR}=(0,v_c,0)$ where $v_c\approx 220\,\text{km/s}$ \cite{Kerr:1986hz} is the local circular speed.
Furthermore, the latest determination of the Sun's peculiar motion finds $\vec{v}_{\odot, \text{pec}} = (11.1, 12.24, 7.25)\,\text{km/s}$ \cite{Schoenrich2010}.
        \begin{figure}[H]
        \begin{minipage}[H]{0.5\linewidth}
        \includegraphics[width=\linewidth]{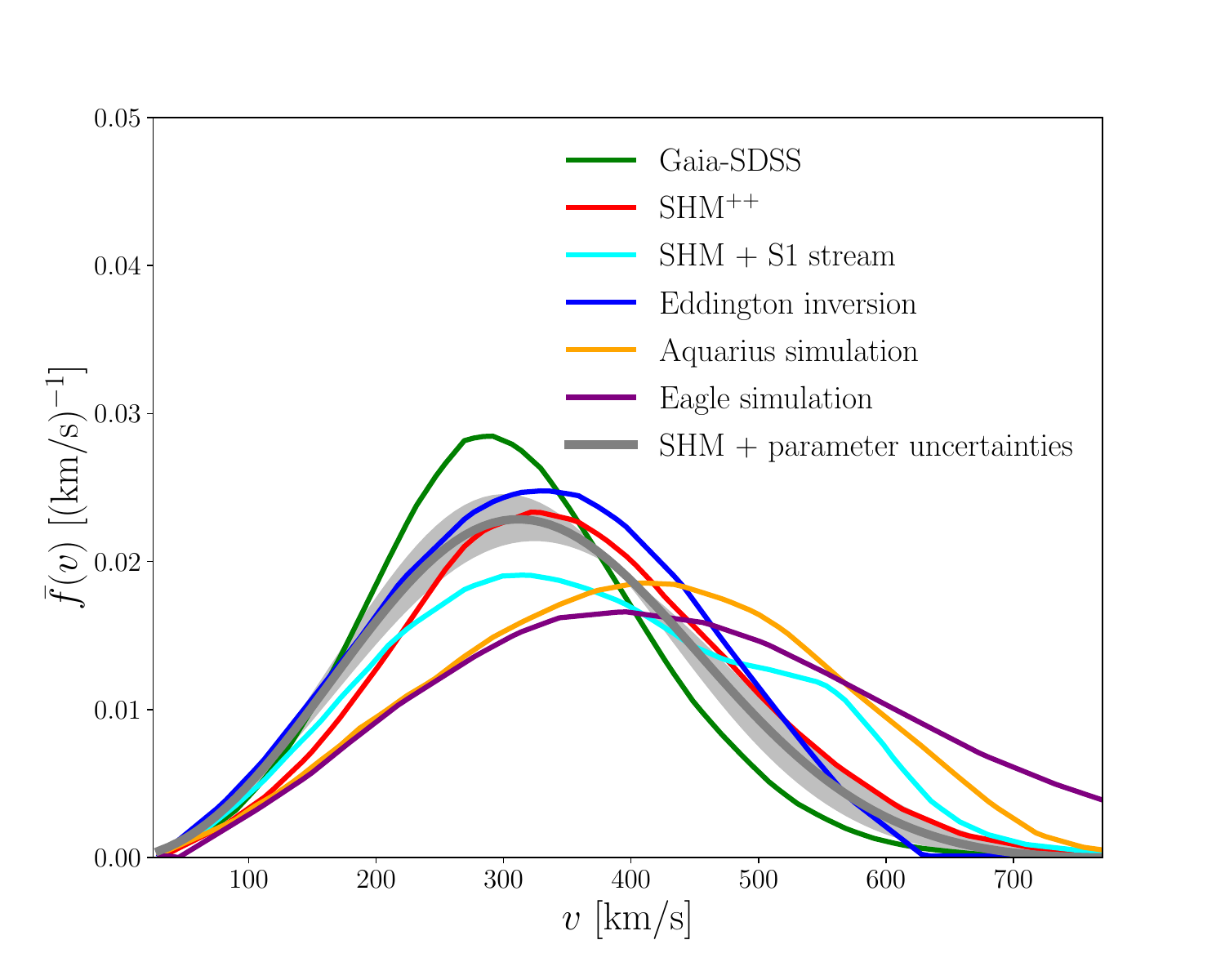}
        \end{minipage}
        \hfill
        \begin{minipage}[H]{0.5\linewidth}
        \includegraphics[width=\linewidth]{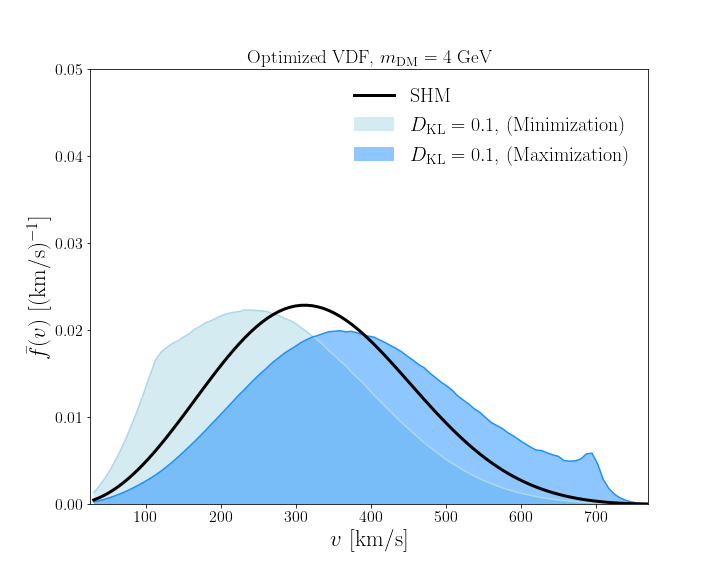}
        \end{minipage}%
    		\caption{\textit{Left plot:} A variety of dark matter velocity distributions (in the Laboratory frame) discussed in the literature and obtained with different methods. \textit{Right plot:} Optimized velocity distributions of a direct detection experiment, for a given dark matter mass and maximal distance that the true distribution can deviate from the Maxwell-Boltzmann (See Section \ref{sec:Nuclear} for details on the optimization problem yielding the aforementioned velocity distributions). The colored regions represent the family of velocity distributions that deviate at most a value of $D_{KL}$=0.1 from the Maxwell-Boltzmann. It is clearly visible that the uncertainties in the dark matter velocity distribution can be very well parameterized with our method.}
    		\label{fig:VDF}
    	\end{figure}
    	\begin{table}[H]
    		\begin{center}
    			\begin{tabular}{l|ccccc}
    				\toprule
    				\toprule
    				& $L^1$ & $L^2$ & $L^\infty$ & $D_{\chi^2}$ & $D_{KL}$\\
    				\bottomrule
    				\toprule
    				SHM parameter uncertainties & $\leq 0.101$ & $\leq 0.024$ & $\leq 0.004$ & $\leq 1.822$ & $\leq 0.040$ \\
    				Eagle simulation \cite{Schaye_2014} &  0.258 & 0.057 & 0.0086 & 9.73 & 0.217 \\
    				Aquarius simulation \cite{Vogelsberger:2008qb} & 0.225 & 0.052 & 0.0083 & 0.730 & 0.144 \\
    				Eddington inversion \cite{Mandal:2018efq} & 0.060  & 0.016 & 0.0034 & 0.030 & 0.018  \\
    				SHM + S1 stream \cite{O_Hare_2018} & 0.118 & 0.029 & 0.0064  & 0.223  & 0.050 \\
    				SHM$^{++}$ \cite{evans2018shm} & 0.067 & 0.016 & 0.0032 & 0.096 & 0.019 \\
    				Gaia-SDSS \cite{Necib:2018iwb} & 0.103   & 0.027  & 0.0064 & 0.060 & 0.037 \\
   				    \bottomrule\\
           
    				Suggested ranges (this work)  & [0.1,1]   & [0.01,0.1]  & [0.001,0.01] & [0.001,10] & [0.01,0.1] \\
    				\bottomrule
    				\bottomrule
    			\end{tabular}
    		\end{center}
    		\caption{Deviations from SHM for different velocity distributions and distance measures. Concretely, we show the distance of the SHM to velocity distributions extracted from the Eagle simulation  \cite{Schaye_2014}, from the Aquarius simulation \cite{Vogelsberger:2008qb}, from Eddington's result \cite{Mandal:2018efq} and from local tracers of dark matter \cite{O_Hare_2018, evans2018shm, Necib:2018iwb}. Furthermore, we show the distance implied by parametric uncertainties of the SHM. We restrict the integration domain by the minimum escape velocity $v_{\rm esc}$ among the two distributions whose distance is determined. Finally, we suggest sensible ranges for each distance measure, as inferred from comparing the Maxwell-Boltzmann with a set of different velocity distributions. Although these ranges are suited for such set of velocity distributions, it should be noted that we can't preclude larger deviations from the Maxwell Boltzmann w.r.t to the true distribution.}
    		\label{tab:CompNorms}
    	\end{table}
As the parameters $\sigma_v$ and $v_\text{esc}$ describing the SHM are subject uncertainties of about 10\%, we also show the deviations of SHM velocity distributions with parameters $v_0=\sqrt{2}\sigma_v$ within the range $[200, 240]\,\text{km/s}$ and $v_\text{esc}$ within $[500, 600]\,\text{km/s}$ to the SHM with the canonical values of $v_0=220\,\text{km/s}$ and $v_\text{esc}=544\,\text{km/s}$.
\newline
As apparent from the results presented in Tab. \ref{tab:CompNorms}, the conclusions drawn from the comparison of different velocity distributions are independent of the distance. However, the scaling of every distance measure changes. Concretely, we find that velocity distributions extracted from the Aquarius and Eagle simulations \cite{Vogelsberger:2008qb, Bozorgnia:2016ogo} and from local tracers of dark matter \cite{Necib:2018iwb} deviate more from the SHM than justified by uncertainties of the SHM input parameters. Furthermore, we find that the velocity distribution determined in Ref. \cite{Mandal:2018efq} using Eddington's result deviates less from the canonical SHM than the deviations due to uncertainties of the SHM parameters.
However, we note that this does not imply the velocity distribution from \cite{Mandal:2018efq} is well fit by the SHM.
    
\section{Upper limits from nuclear recoil searches}\label{sec:Nuclear}
	
The knowledge of the magnitude of the distance to the Maxwell-Boltzmann for various velocity distributions can be used to quantify the impact of uncertainties on the results of direct dark matter searches.
Concretely, we will derive upper limits from the CRESST-III, LUX-ZEPLIN and PICO-60 experiments \cite{Abdelhameed:2019hmk,LUX-ZEPLIN:2022qhg,PICO:2019vsc} which aim to detect dark matter by counting the number of scattering events between dark matter particles and nuclei in a detector.
The expected number of events can be calculated by multiplying the exposure of the experiment by the recoil rate $R$ that is given by \cite{Cerdeno:2010jj}
    \begin{align}
	R = \int_{0}^{\infty}\,\sum\limits_i \epsilon_i(E_R)\,\frac{\text{d}R_i}{\text{d} E_R}\,\text{d} E_R\,.\label{eq:RecoilRate}
	\end{align}
The recoil spectrum of dark matter particles off the nuclear species $N_i$ is given by
	\begin{align}
	\frac{\text{d} R_i}{dE_R}=\frac{\xi_i\;\rho_\text{loc}}{\mdm\;m_{N_i}}\int_{v^\geq v_\text{min,i}(E_{R})} \text{d}^3\vec{v} \, v\,f(\vec{v})\,\frac{\text{d}\sigma_i}{dE_R} (v,E_R)\,,
	\end{align}
where $\vec{v}$ is the velocity of the dark matter particle in the laboratory frame \cite{McCabe:2013kea}. We use the software \texttt{DDCalc} to calculate the nuclear recoil rates in CRESST-III, PICO60 and LUX-ZEPLIN \cite{GAMBITDarkMatterWorkgroup:2017fax, GAMBIT:2018eea}. The differential scattering cross section of a dark matter particle with a specific nucleus $N_i$ with mass $m_{N_i}$ and mass fraction $\xi_i$ inside the detector is denoted by $\text{d}\sigma_i/\text{d}E_{R}$.
A dark matter particle must have a velocity larger than $v_{\text{min},i}(E_R) = \sqrt{m_{N_i} E_{R}/(2\mu_{N_i}^2)}$ in order to transfer energy $E_{R}$ onto the nucleus, where $\mu_{N_i}$ is the reduced mass of the dark matter-nucleus system. We model the dark matter velocity distribution as a superposition of 500 streams\footnote{ Here streams corresponds to Dirac deltas, \textit{i.e} with zero velocity dispersion. In reality, dark matter streams should have some small but finite velocity dispersion, analogously to well known Galactic stellar streams.} \cite{Ibarra:2017mzt}
	\begin{align}
	f(\vec{v}) = \sum\limits_j\,f(\vec{v}_j)\,w_j\,\delta(\vec{v}-\vec{v}_j),
    \label{eq:discretization_vdf}
	\end{align}
where $w_i$ are integration weights from discretizing the integration and $f(\vec{v}_j)$ is the velocity distribution evaluated at the velocity $\vec{v}_j$\footnote{The decomposition into a finite number of streams is justified by the Fenchel-Eggleston theorem \cite{Fenchel1953ConvexCS} and Choquet's theorem \cite{SB_1956-1958__4__33_0}.}.
Defining $x_j\equiv f(\vec{v}_j)\,w_j$, we can relate the numerical integration to a decomposition of the velocity distribution into streams.
Applying this to the recoil rate at a direct detection experiment yields
	\begin{align}
	R =  \sum\limits_{j} R(\vec{v}_j)\,x_j\,,
    \label{eq:discretization_rate}
	\end{align}
where $R(\vec{v}_j)$ is the response of the detector to a stream of dark matter particles with velocity $\vec{v}_j$.
This can be calculated from Eq. \eqref{eq:RecoilRate} by replacing the velocity distribution with a delta function
	\begin{equation}
	\label{eq:DD_Response}
	R(\vec{v}_j)=\sum\limits_i\,\int_0^\infty\dER\,\frac{\xi_i\;\rho_\text{loc}}{\mdm\;m_{N_i}}\Theta\left(v_j- v_\text{min,i}(E_{R})\right)\, v_j\,\frac{\text{d}\sigma_i}{dE_R} (v_j,E_R)\,.
	\end{equation}
This discretization allows us to apply the optimization techniques discussed in section \ref{sec:FuncOpt1}. We aim to optimize the number of recoil events induced by dark matter in an experiment, subject to the constraint that the dark matter velocity distribution deviates at most a given value $\Delta$ from a Maxwell-Boltzmann distribution, that it is positively defined, and that all dark matter particles reaching the Earth are bound to the Galaxy. The optimization over a domain of velocities $\vec{v}_{0}$ reads:

\begin{center}
\begin{equation}\label{eq:optproblemcontinuousdivergences}
    \text{\textbf{Optimize:} } \mathcal{R}[f]=\int d^{3}v_0 f(\vec{v}_{0})R_{\vec{v}_{0}}
\end{equation}
\textbf{subject to:}
\begin{equation*}
\int d^{3}v_{0} f(\vec{v}_{0})=1
\end{equation*}
\begin{equation*}
D(f_{\rm MB},f) \leq \Delta
\end{equation*}
\begin{equation*}
f(\vec{v}_{0}) \geq 0
\end{equation*}
\begin{equation*}
v_{0} \leq v_{\rm esc}
\end{equation*}
\end{center}

where the constraint $D(f,f_{\rm MB}) \leq \Delta$ imposes that the maximal distance between the true dark matter distribution and the Maxwell-Boltzmann is given by $\Delta$, where the devitation is parametrized with a distance measure as those described in section \ref{sec:Distances}. This problem can be formulated on general grounds, as we have done, but one needs to write it in its discretized form for each distance measure under consideration and check the convexity of the problem. In the following, we will focus on two norms ($L^1$ and $L^{\infty}$) and two information divergences (KL-divergence and $\chi^{2}$ divergence). These $L^1$-norm, $L^{\infty}$-norm and the $\chi^{2}$-divergence satisfy the convexity requirements of Problem \ref{eq:optproblemcontinuousdivergences} and can be recasted as Second Order Cone problems (SOCP's)\footnote{Second Order Cone problems refer to problems with constraints lying in the second order cone in $\mathbb{R}^{n+1}$, defined as $\mathcal{C}_{n+1}=\left\{\left.\left[\begin{array}{l}
x \\
t
\end{array}\right] \right\rvert\, x \in \mathbb{R}^n, t \in \mathbb{R},\|x\|_2 \leq t\right\}$}, solvable by \texttt{CVXPY}, see \cite{Rappelt:2020tch} for further details.
\clearpage
	
\begin{figure}[H]
\begin{minipage}[H]{0.45\linewidth}
\includegraphics[width=\linewidth]{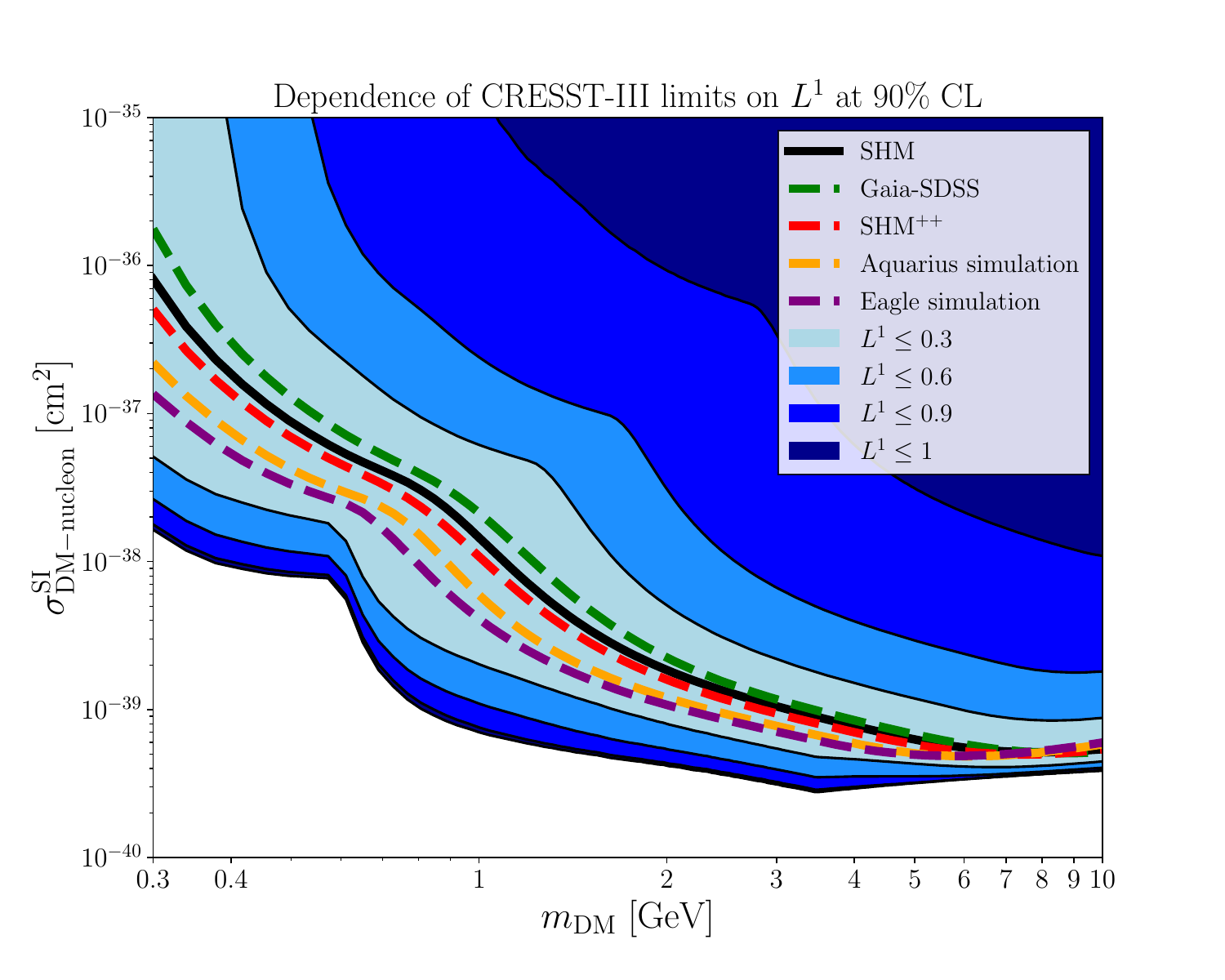}
\end{minipage}
\vspace{-4mm}
\begin{minipage}[H]{0.45\linewidth}
\includegraphics[width=\linewidth]{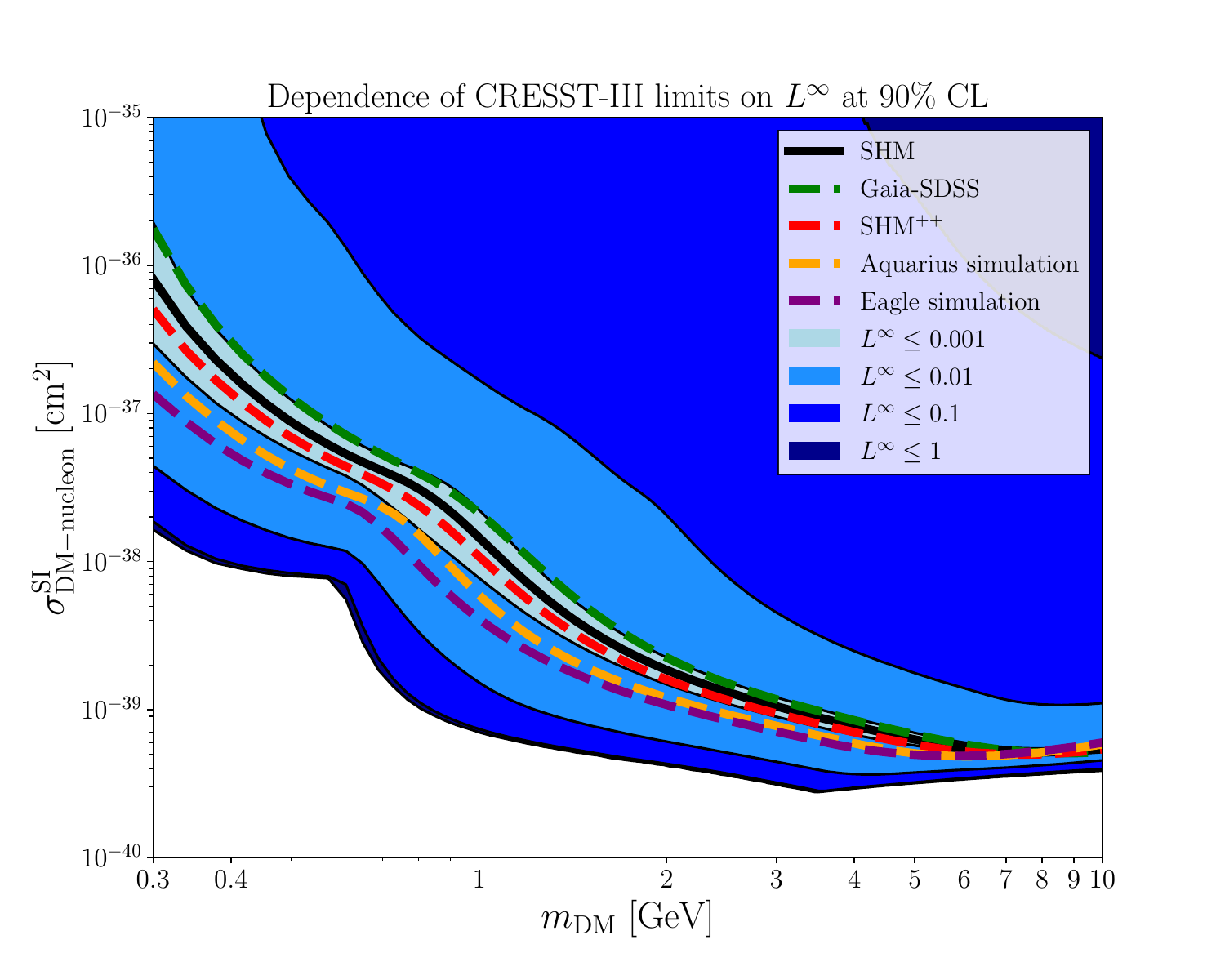}
\end{minipage}%
\end{figure}
\vspace{-4mm}
\begin{figure}[H]
\begin{minipage}[H]{0.45\linewidth}
\includegraphics[width=\linewidth]{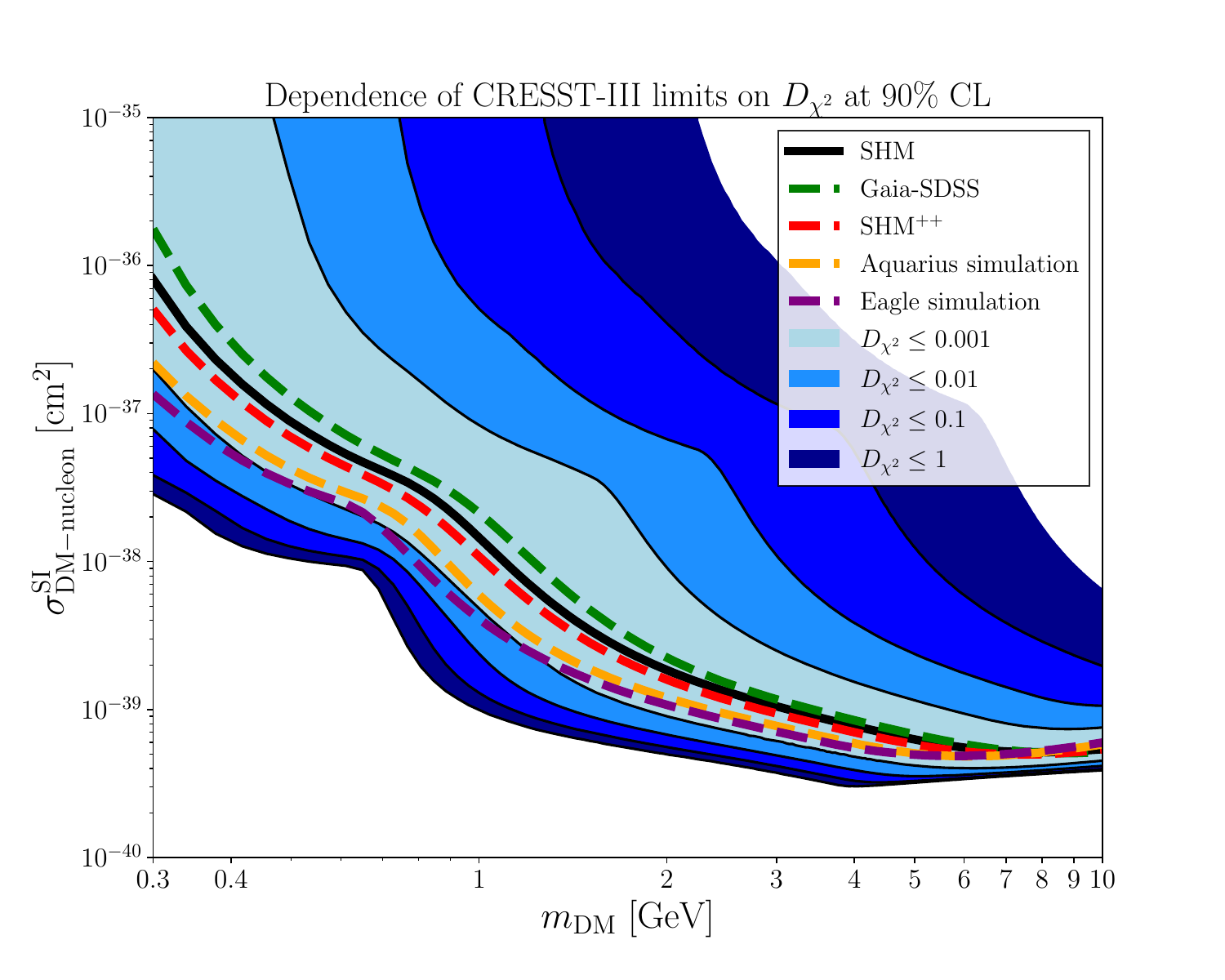}
\end{minipage}
\vspace{-4mm}
\begin{minipage}[H]{0.45\linewidth}
\includegraphics[width=\linewidth]{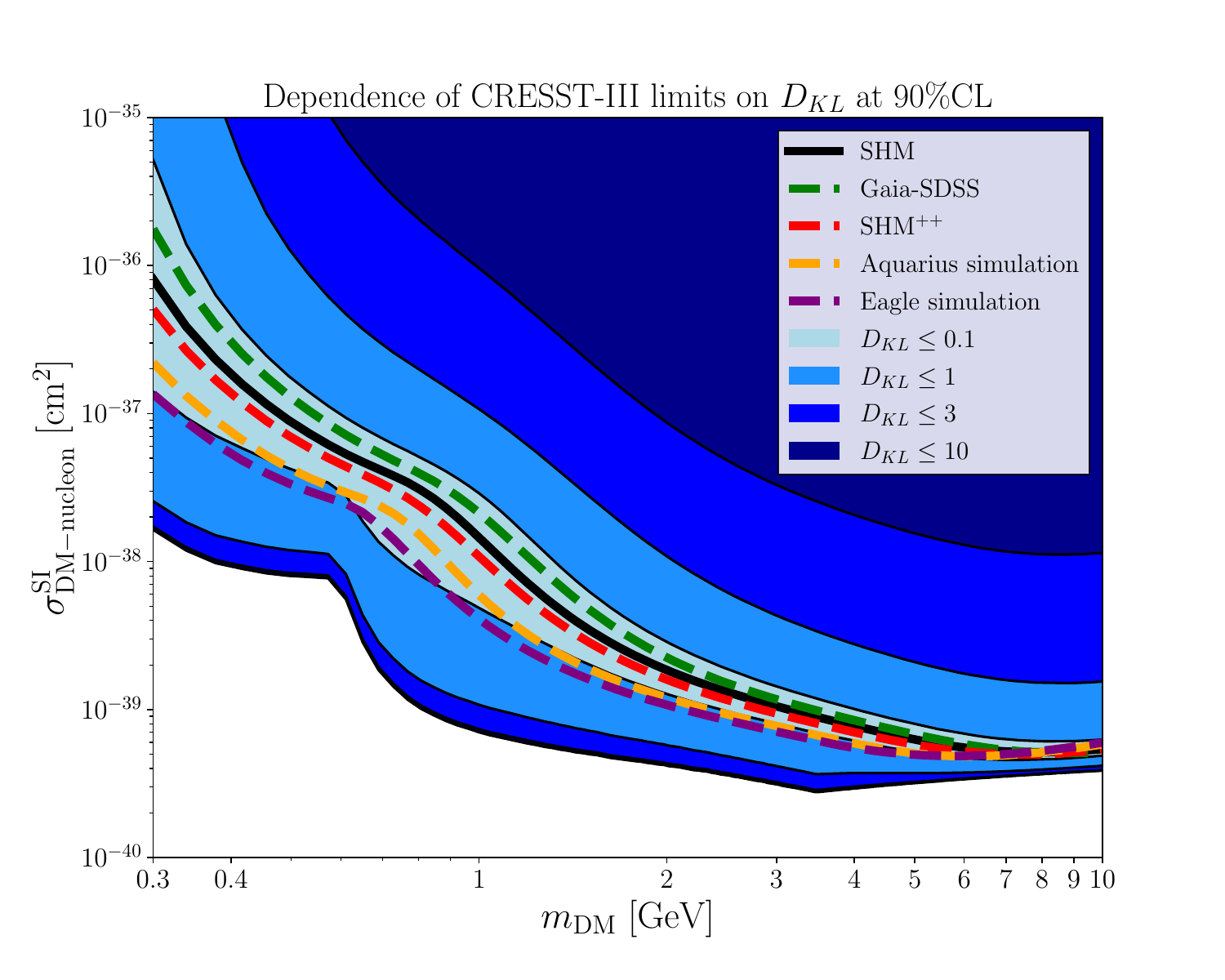}
\end{minipage}
\vspace{-4mm}
\begin{figure}[H]
\begin{minipage}[H]{0.45\linewidth}
\includegraphics[width=\linewidth]{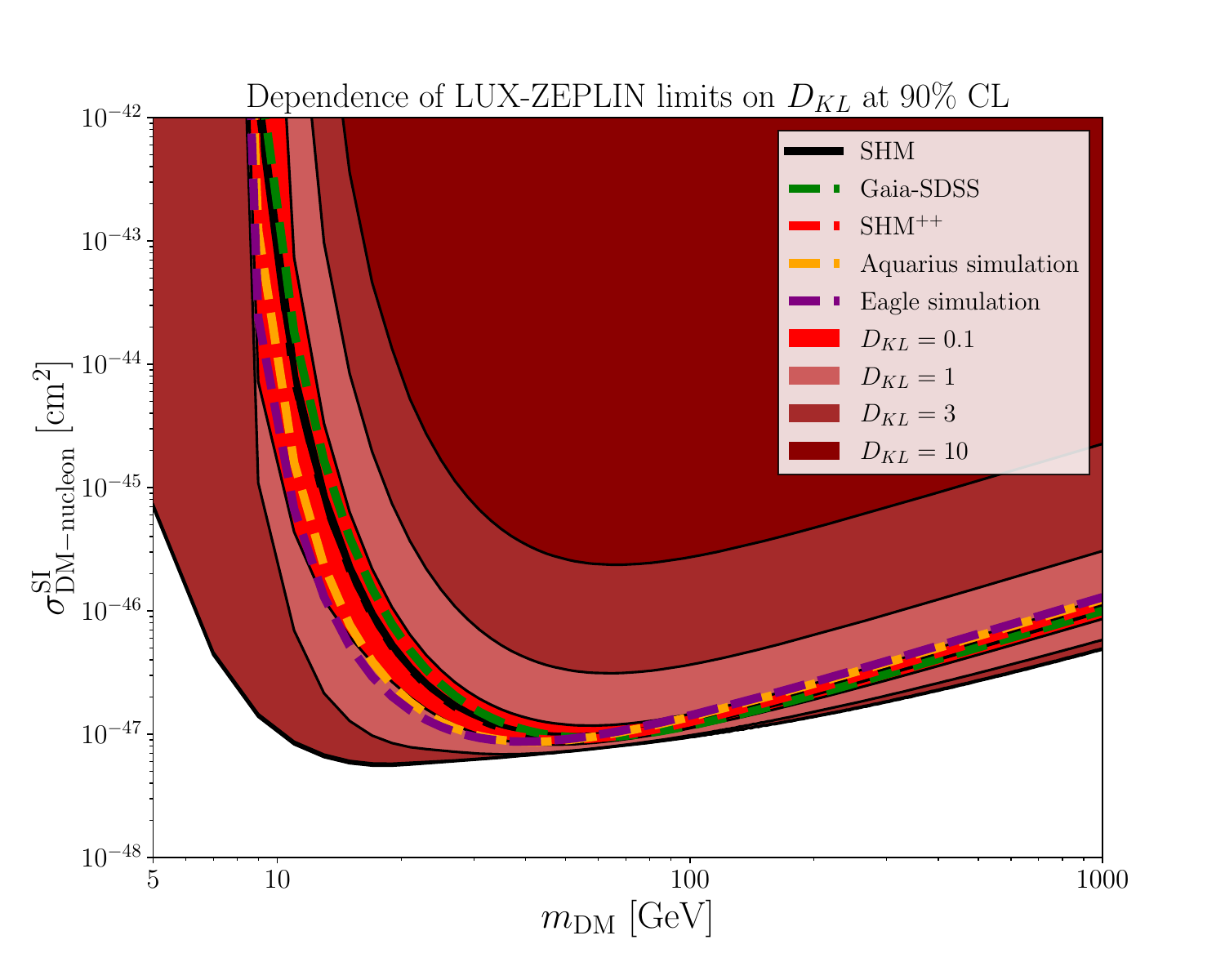}
\end{minipage}%
\vspace{-4mm}
\begin{minipage}[H]{0.45\linewidth}
\includegraphics[width=\linewidth]{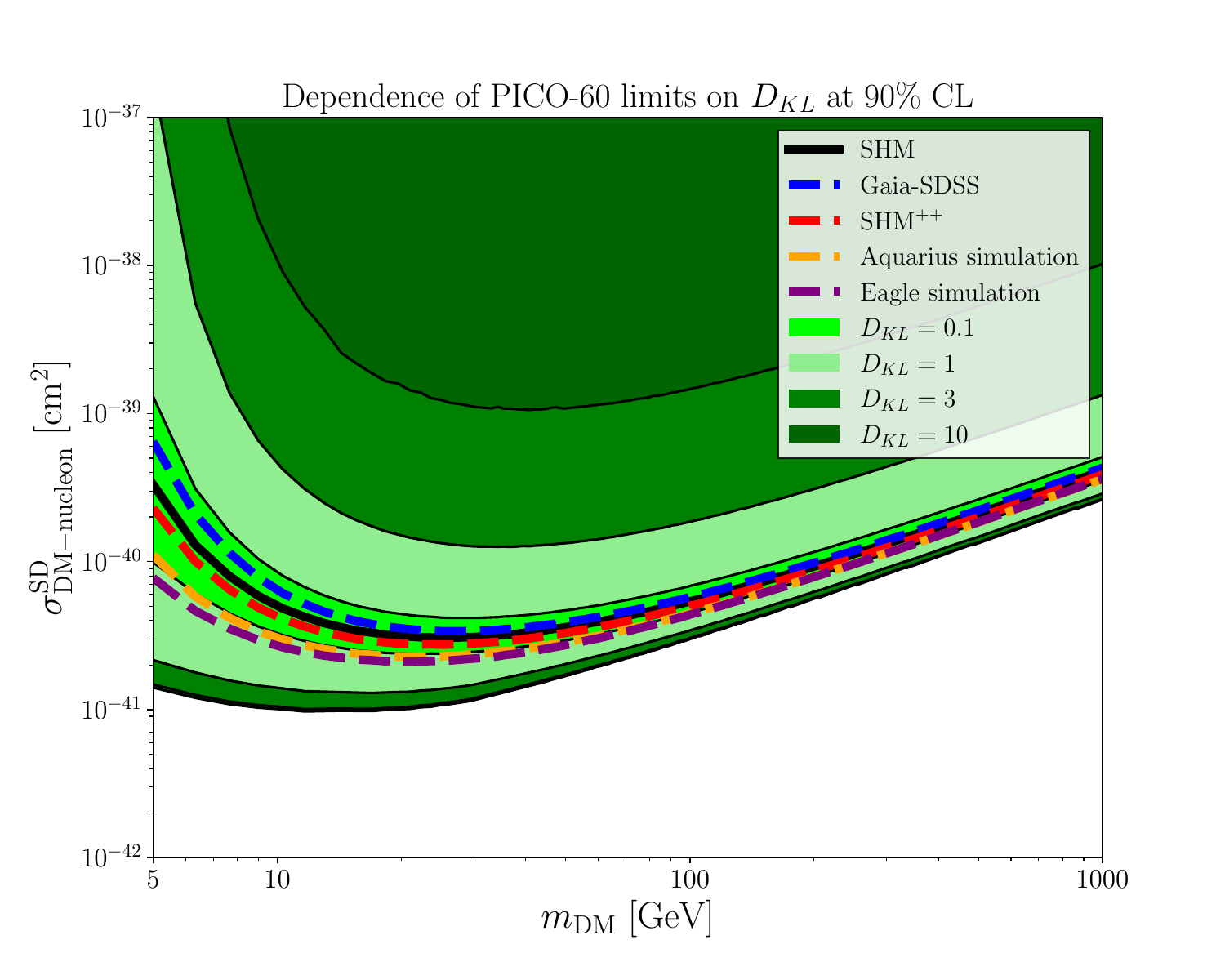}
\end{minipage}%
\end{figure}
\caption{\textit{Upper and middle panels}: 90\% CL upper limits on the spin-independent dark matter-nucleon cross section from the CRESST-III experiment, for different deviations w.r.t the SHM. The coloured bands represent the set of upper limits obtained from those feasible velocity distributions that satisfy the experimental constraints and deviate at most a certain value with respect to the SHM. From upper left to lower right, we show the results obtained for the $L^{1}$-norm, $L^{\infty}$-norm, Neyman $\chi^{2}$-divergence and KL-divergence. For comparison, we show in colored dotted lines the limits derived for four different halo models. \textit{Lower panels:} Constraints on the spin-independent dark matter-nucleon cross section from the LUX-ZEPLIN experiment, and on the spin-dependent dark matter-nucleon cross section from the PICO-60 experiment, for different deviations w.r.t the SHM parametrized by the KL-divergence.}
\label{fig:cresstiiisi}
\end{figure}

\clearpage

The KL-divergence can also be converted into a convex constraint by transforming it into a form in which it belongs to the exponential cone. The convex constraint would be given by $\Delta \geq D_{KL}(f_{\rm MB},f)$ and is equivalent to
\vspace{4mm}

\begin{equation}
 \Delta \geq D(f_{\rm MB}, f) \Longleftrightarrow-\Delta \leq f_{\rm MB} \log (f / f_{\rm MB}) \Longleftrightarrow(f, f_{\rm MB},-\Delta) \in K_{\exp },
\end{equation}
\vspace{3mm}

where the exponential cone $K_{\exp }$ is described in Appendix \ref{sec:A2}.  We would like to stress that in the literature the KL-divergence is typically used in convex optimization problems as the objective function, but not as a constraint of the optimization problem under consideration. Our methodology is perhaps not new but definitely not widespread, not only in this particular physics context but in generic convex optimization studies.\\

We can apply the developed methodology to derive upper limits in a direct detection experiment for different values of the distance measure parametrizing the deviation from the Maxwell-Boltzmann velocity distribution. In Fig. \ref{fig:cresstiiisi} we present optimized upper limits on the dark matter mass-cross section parameter space at $90\%$ C.L from the CRESST-III, LUX-ZEPLIN and PICO-60 experiments. In the following, we discuss and interpret the results in some detail.
	
When solving the problem stated in Eq. \ref{eq:optproblemcontinuousdivergences} by maximizing the number of events with a sufficiently large maximal deviation between the Maxwell-Boltzmann distribution and the true distribution, the optimized velocity distributions approach a single stream of dark matter particles moving at the Milky Way escape velocity, and all distance measures perform similarly. When minimizing, on the other hand, the discrepancy between different distances is not only caused by a different scaling of each distance, but also due to the way in which the distance measure applies. At low masses, the $L^{1}$-norm behavior can be explained kinematically. As its magnitude increases, the high velocity tail is suppressed and, at a certain deviation, the experiment is not sensitive to such velocity distributions anymore. The $\chi^{2}$-divergence shows a similar behavior as the $L^{1}$-norm, being the deviation at low masses ($m_{\rm DM} \sim$  0.3 GeV) very large, and compensated at higher masses ($m_{\rm DM} \sim$ 10 GeV), where the experiment is kinematically sensitive to most velocity spectrum.
	
The $L^{\infty}$-norm and the KL-divergence perform quite differently. The $L^{\infty}$-norm parametrizes a larger impact of the velocity distribution at low masses, but the optimized velocity distributions do not lie completely below the velocity threshold of the experiment. They rather retain a fraction of the kinematically accessible spectrum, such that optimized upper limits do not deviate so sharply at low masses from the SHM one as it happens for other distance measures. 

\begin{figure}[H]
\includegraphics[width=0.49\linewidth]{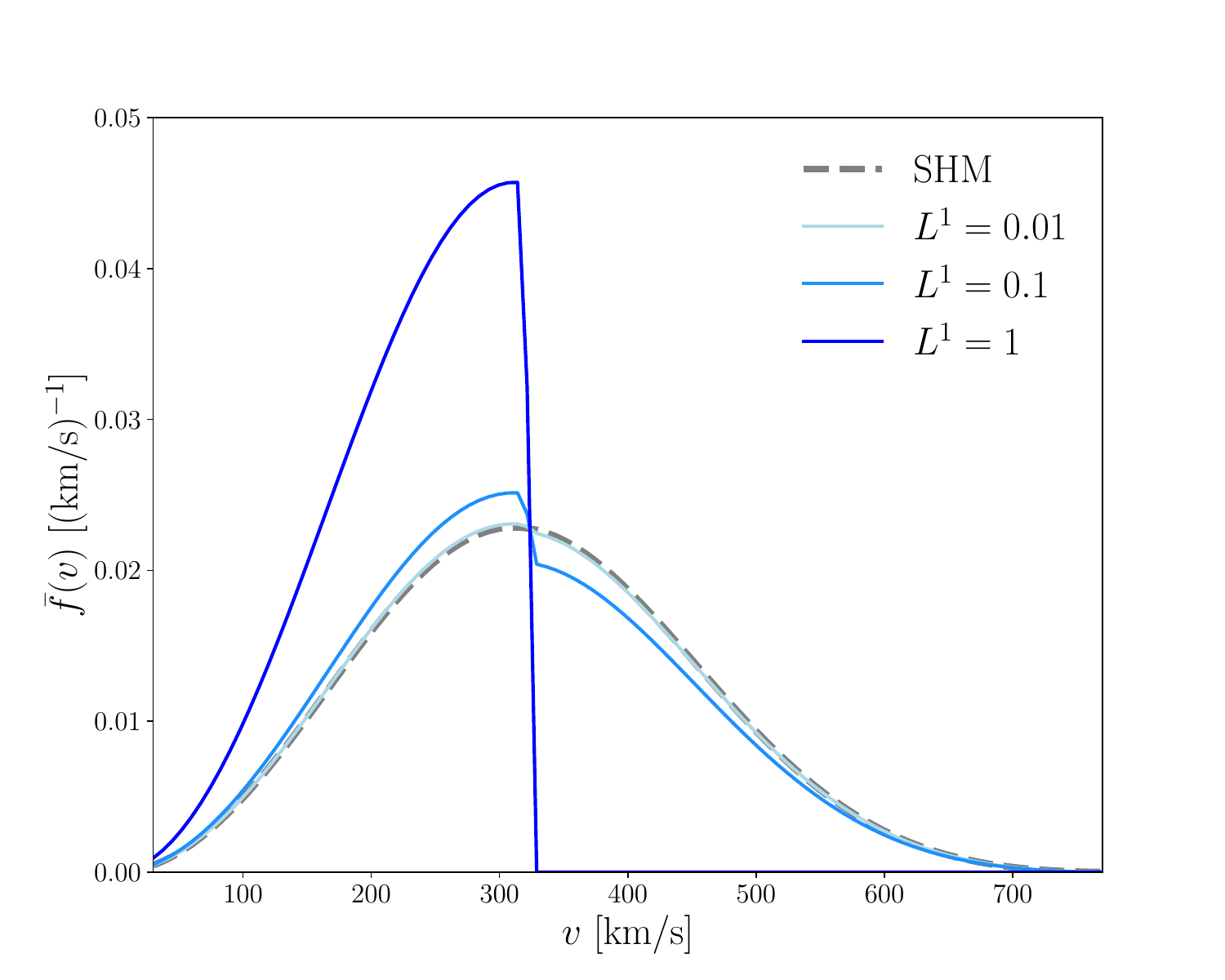}
\includegraphics[width=0.49\linewidth]{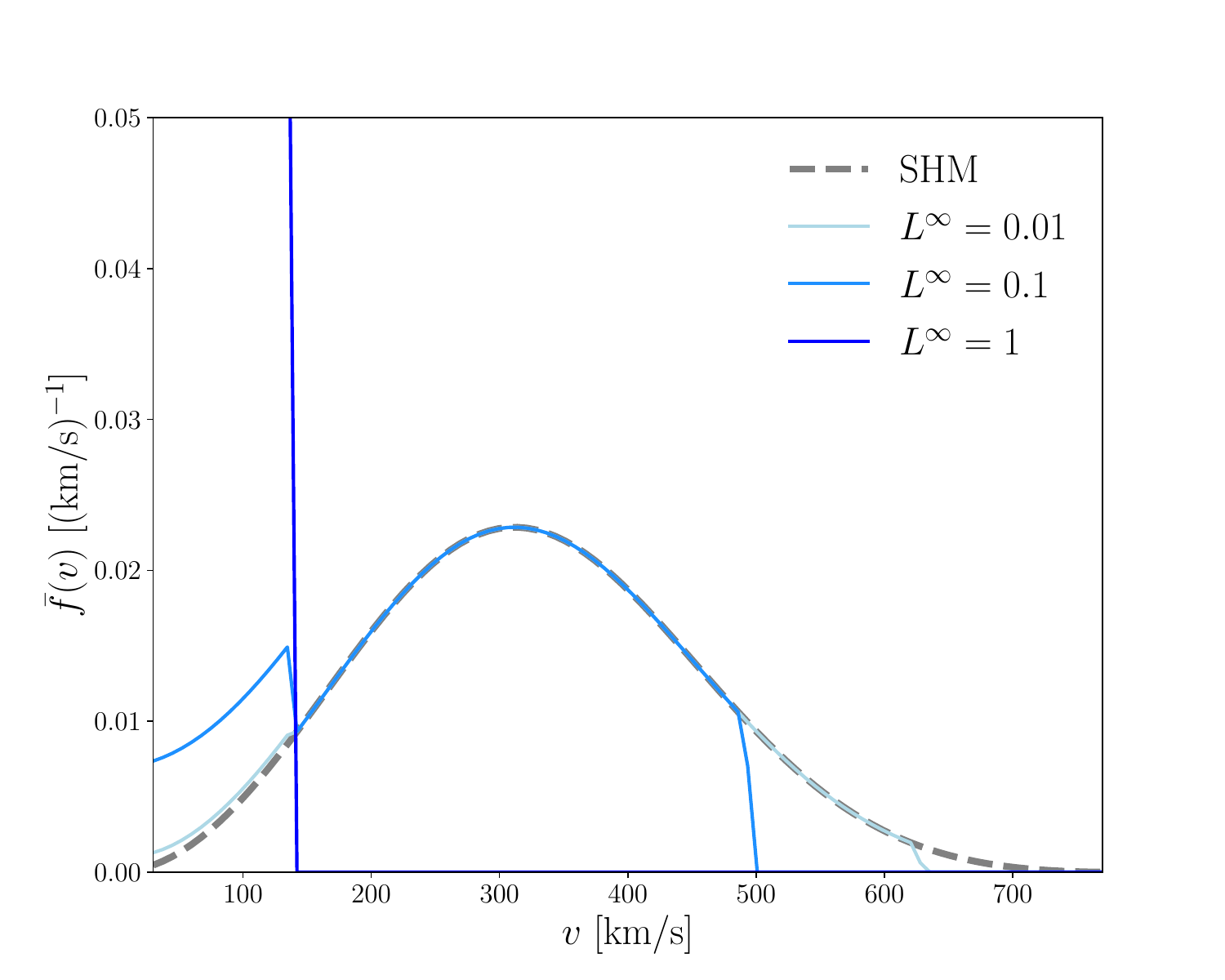}
\includegraphics[width=0.49\linewidth]{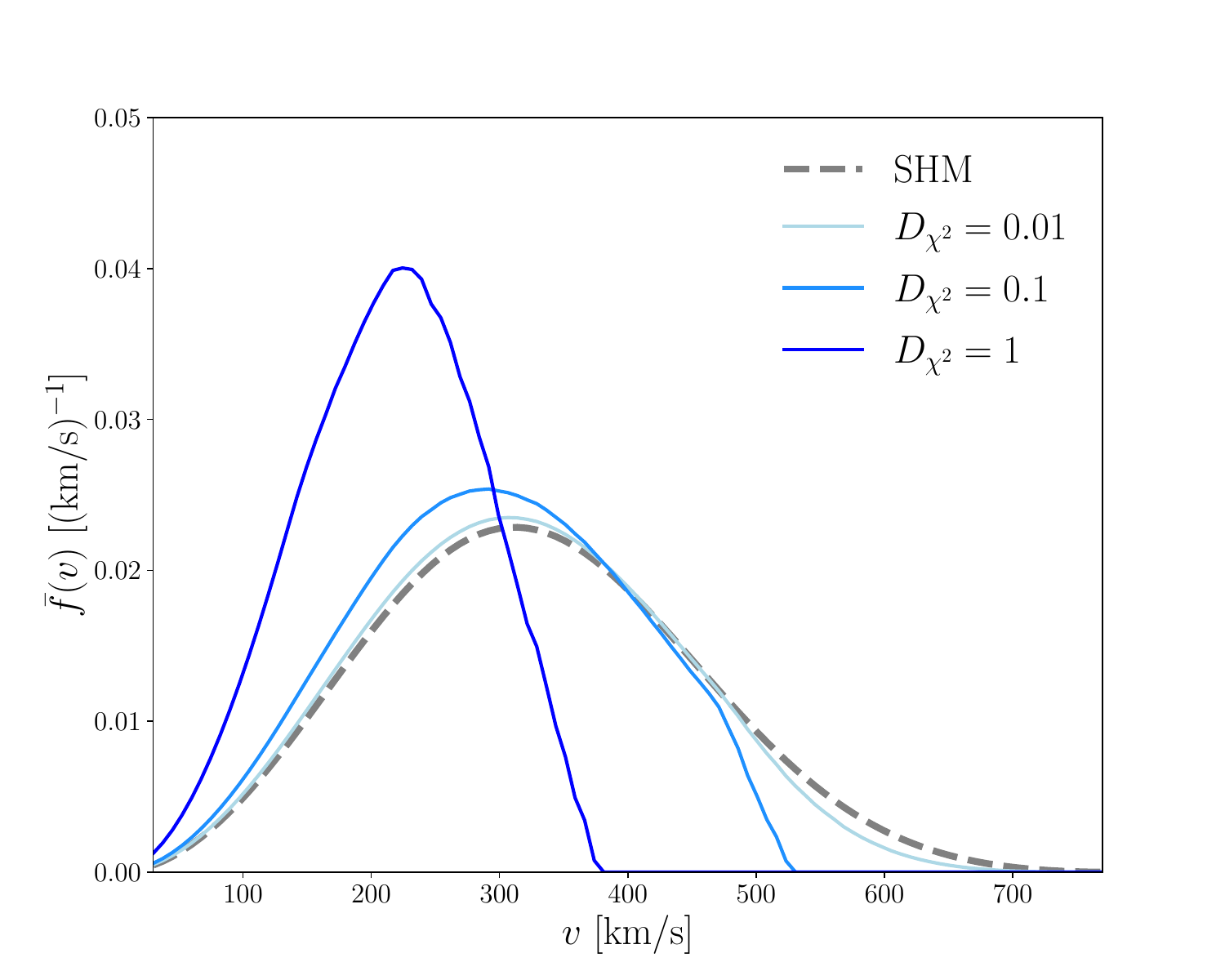}
\includegraphics[width=0.49\linewidth]{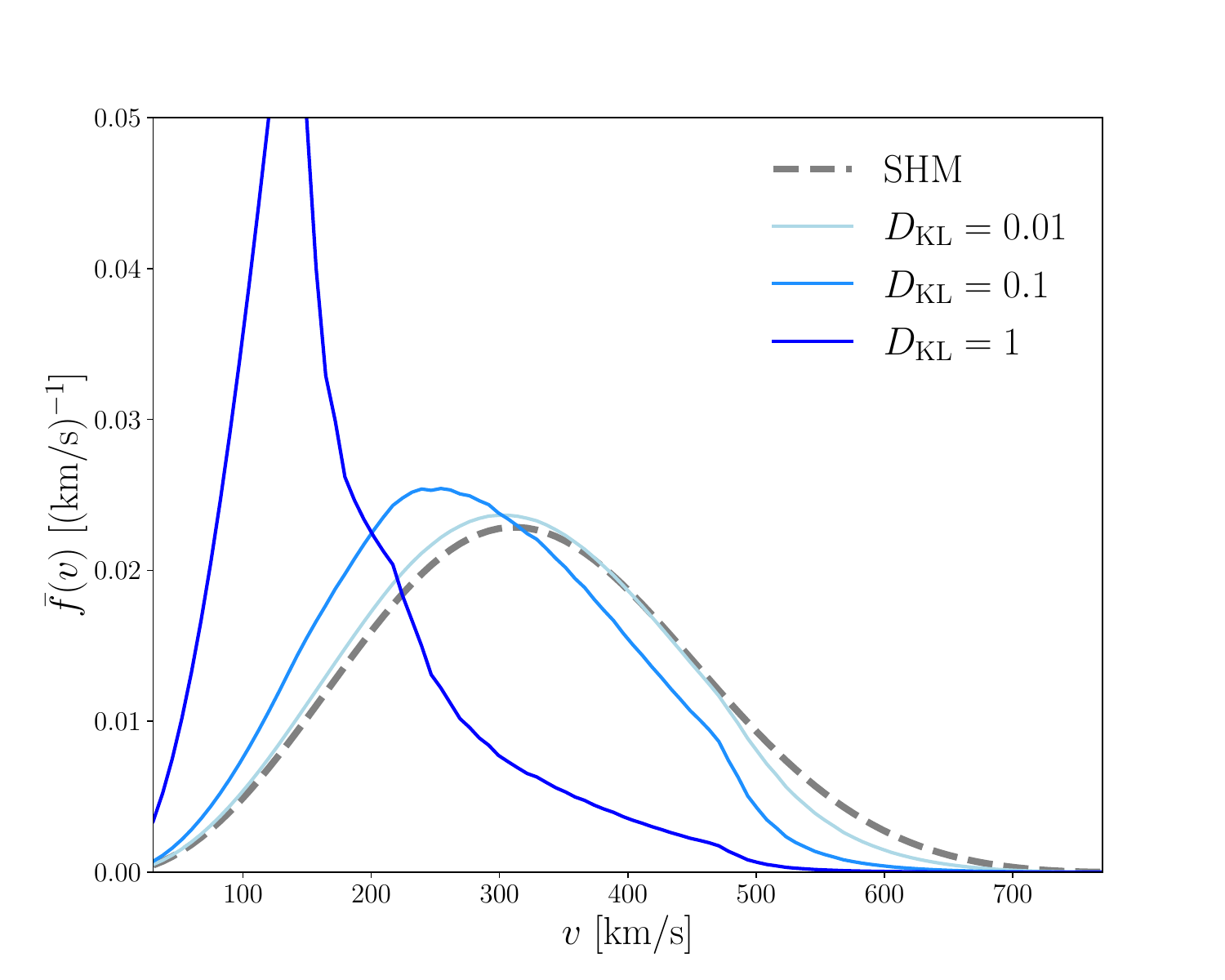}
\label{fig:VDFs}
\caption{Optimized velocity distributions of Problem \ref{eq:optproblemcontinuousdivergences}, for different distance measures: $L^1$ norm in the upper left, $L^{\infty}$ in upper right, $D_{\chi^2}$ in lower left, and $D_{\rm KL}$ in lower right. For each distance measure, different maximum values of the deviation from the SHM are considered (see main text for details). The dark matter mass is fixed at $m_{\rm DM}$, and the experiment considered is CRESST-III. For comparison, we also show in dotted grey color the SHM Maxwell Boltzmann velocity distribution.}
\label{fig:optimized_vdf}
\end{figure}

This can be appreciated in Figure \ref{fig:optimized_vdf}, where we show optimized velocity distributions for different distance measures and values of $\Delta$. This effect is even more pronounced in the case of the KL-divergence, which does not suppress the high velocity tail at low dark matter masses. This explains that the upper limits do not deviate so largely with respect to the SHM upper limit.

The KL-divergence parametrization suggests, against our kinematic intuition, that there could be feasible velocity distributions that cause an uncertainty in the results shown by a given experiment that is larger at high masses than at low masses. This can happen since the KL-divergence penalizes deviations at the high velocity tail strongly, thus smoothly interpolates between the Maxwell-Boltzmann-like velocity distribution and the most aggressive/conservative velocity distributions. For example, the CRESST experiment is sensitive to the complete dark matter velocity spectrum at 10 GeV, hence, differences between the distributions at low masses are taken into account. For sufficiently low masses ($m_{\rm DM} \leq 1$ GeV), only the high velocity tail is seen and there the KL does not penalize it so strongly, therefore being the difference in the upper limits not so large. We notice that the inequality Eq. \ref{eq:ineq1} holds for our analysis. Indeed, the $L^{\infty}$-norm scaling is significantly lower than the KL-divergence, obtaining similar deviations for values of $L^{\infty}$= 0.001 and $D_{KL}$ = 0.1. We have, as well, shown the upper limits for four specific halo models: from the Aquarius and Eagle simulations, and the SHM$^{++}$ and Gaia-SDSS combined data velocity distributions. We notice that for each distance measure, even though the deviation is not so strong, $\mathcal{O}(1)$, these fall in different bands, which motivates the need of using different distance measures to make a proper halo-independent analysis. As we have already mentioned, in Tab. \ref{tab:CompNorms}, we have computed the distance measure between different halo models and the Maxwell-Boltzmann velocity distribution, in order to get a hint of which deviations of our analysis are consistent with experimental results and are physically motivated. We notice that, in the case of tracers studies, we obtain deviations in the two closest bands to the SHM upper limit, leading in some regions to uncertainties of $\mathcal{O}(10)$. For the case of N-body simulations, these uncertainties can reach values of $\mathcal{O}(10^{2})$ in some regions of the parameter space. These of course depend on the distance measure under consideration and we remark that for the KL-divergence we do not encounter physically motivated deviations larger than $\mathcal{O}(1)$.\\

\section{Upper limits from electron recoil searches}\label{sec:Electron}
	We can also interpret the results of dark matter-electron scattering searches using the method presented in sections \ref{sec:FuncOpt1} and \ref{sec:Distances}. The results may vary w.r.t the nuclear recoil case due to different kinematics (the recoiling electron is bound without fixed momentum), and different energy threshold from experiments. In liquid xenon, for a given stream velocity $\vec{v}$ in the galactic frame with modulus $v$, the differential ionization cross section reads \cite{Essig:2011nj}
\begin{equation}
    \frac{d\sigma_{ion}^{nl }}{d\text{ln}  E_{er}}(v,E_{er})=\frac{\bar{\sigma}_{e}}{8 \mu_{e}^{2}v^{2}}\int_{q^{nl}_{\rm min}}^{q^{nl}_{\rm max}} dq q \left |f_{ion}^{nl}(k',q)  \right |^{2} \left |F_{\mathrm{DM}}(q)  \right |^{2}.
\end{equation}
where the minimum and maximum momentum transfer in the orbital $(n,l)$ is given by
\begin{equation}
    q^{nl}_{\substack{{\rm max}\\{\rm min}}}(E_{er})= m_{\rm DM} v \pm \sqrt{m_{\rm DM}^{2}v^2-2m_{\rm DM}(E_{er}+\left |E^{n,l}  \right |)}
    \label{eq:q_max_min}
\end{equation}
and the differential ionization rate is
\begin{equation}
\frac{dR_{ion}^{\vec{v}}}{d\text{ln}E_{er}}= N_{T}\frac{\rho_{\rm DM}}{m_{\rm DM}}\frac{1}{v} \, \sum_{n,l}\frac{\text{d} \sigma_{\rm ion}^{nl}}{d\text{ln}E_{er}}(v, E_{er}) \,,
\end{equation}\label{eq:diff_ionization_rate_xe}
Here, $\mu_{\rm DM,e}$ is the reduced mass of the dark matter-electron system, 
$\bar{\sigma}_{\rm DM-e}$ is the dark matter-free electron scattering cross section at fixed momentum transfer $q=\alpha m_{e}$, $\left|f_{ion}^{nl}(k',q)  \right |^{2}$ is the ionization form factor of an electron in the $(n,l)$ shell with final momentum $k'=\sqrt{2m_{e}E_{er}}$ and momentum transfer $q$, and $F_{\mathrm{DM}}(q)$ is a form factor that encodes the $q$-dependence of the squared matrix element for dark matter-electron scattering and depends of the mediator under consideration. Finally, the total number of expected ionization events for a given stream reads ${\cal N}=R_{ion}\cdot \mathcal{E}$, with $R_{ion}$ the total ionization rate, calculated from integrating Eq.(\ref{eq:diff_ionization_rate_xe}) over all possible recoil energies, and  $\mathcal{E}$ the exposure ({\it i.e.} mass multiplied by live-time) of the experiment. In our analysis, we consider the ionization of electrons in the following orbitals (with binding energies in eV shown in parenthesis: $5p^{6}$ (12.4), $5s^{2}$ (25.7), $4d^{10}$ (75.6), $4p^{6}$ (163.5) and $3s^{2}$ (1148.7). The corresponding ionization form factors were obtained from the software \texttt{QEDark} \cite{Essig:2015cda}. For the dark matter form factor, we adopt three different parametrizations: the case of a heavy hidden photon $A'$ mediator $m_{A'} \gg q$, with $F_{\mathrm{DM}}(q)$=1, an interaction via an electric dipole moment $F_{\mathrm{DM}}(q)=\alpha m_{e}/q$, and an interaction via an ultralight or massless hidden photon $m_{A'} \ll q$, with $F_{\mathrm{DM}}(q)=\alpha^{2}m_{e}^{2}/q^{2}$.

For semiconductor detectors, the dark matter-electron scattering rate for a given stream velocity \textbf{v} is given by \cite{Knapen:2021run, Knapen_2022}
\begin{equation}
R_{ion}^{\vec{v}}=\frac{1}{\rho_{T}} \frac{\rho_{\rm DM}}{m_{\rm DM}} \frac{\bar{\sigma}_{e}}{\mu_{e}^{2}} \frac{\pi}{\alpha} \int \frac{d^{3} \mathbf{k}}{(2 \pi)^{3}} k^{2}\left|F_{\rm DM}(k)\right|^{2} \int \frac{d \omega}{2 \pi} \frac{1}{1-e^{-\beta \omega}} \operatorname{Im}\left[\frac{-1}{\epsilon(\omega, \mathbf{k})}\right] \delta\left(\omega+\frac{k^{2}}{2 m_{\chi}}-\mathbf{k} \cdot \mathbf{v}\right)
\end{equation}
where $w$ is the energy deposited in the material, $\vec{q}$ is the momentum transfer of the process, and $\rho_T$ is the target density. The rate involves an integration of the Electronic Loss Function (ELF) of the target material, which we calculate with \texttt{DarkELF} \cite{Knapen_2022}. For the dielectric function $\epsilon(\omega, \mathbf{k})$, we use the Lindhard method, which treats the target as a non-interacting Fermi liquid.

In order to calculate the rate for an arbitrary velocity distribution, it is simply necessary to calculate
\begin{equation}
    R_{ion}=\sum_{i} a_{\vec{v}_{i}} R_{ion}^{\vec{v}_{i}}
\end{equation}
where $a_{i}$ is the fraction of dark matter particles moving with velocity $\vec{v_{i}}$.\\

We want to derive 90$\%$ C.L upper limits on the dark matter-electron cross section from SENSEI, when varying the dark matter flux, and given a maximal deviation from the flux determined by the SHM. In this case, unlike we did in our analysis for nuclear recoils in section \ref{sec:Nuclear}, we will allow for the possibility that a component of dark matter particles with velocities larger than the escape velocity of the Milky Way reaches the Earth. Thus, we solve the following optimization problem
\begin{center}
\text{\textbf{Optimize:} }
\begin{equation}\label{eq:optproblemdiscrete}
    \sum_{i} a_{\vec{v}_{i}}R_{ion}^{\vec{v}_{i}}
\end{equation}
\textbf{subject to:}
\begin{equation*}
D_{\rm KL}(f_{\rm MB},f) \leq \Delta
\end{equation*}
\begin{equation*}
\sum_{i=1} a_{\vec{v}_{i}}=1
\end{equation*}
\begin{equation*}            
a_{\vec{v}_{i}} \geq 0
\end{equation*}
\end{center}
where the relation between $f$ and the discretized coefficients $a_i$ is analogous here to our previous discussion for nuclear recoils, explicit in Eq. \ref{eq:discretization_vdf} and Eq. \ref{eq:discretization_rate}.

\begin{figure}[H]
\includegraphics[width=0.5\textwidth]{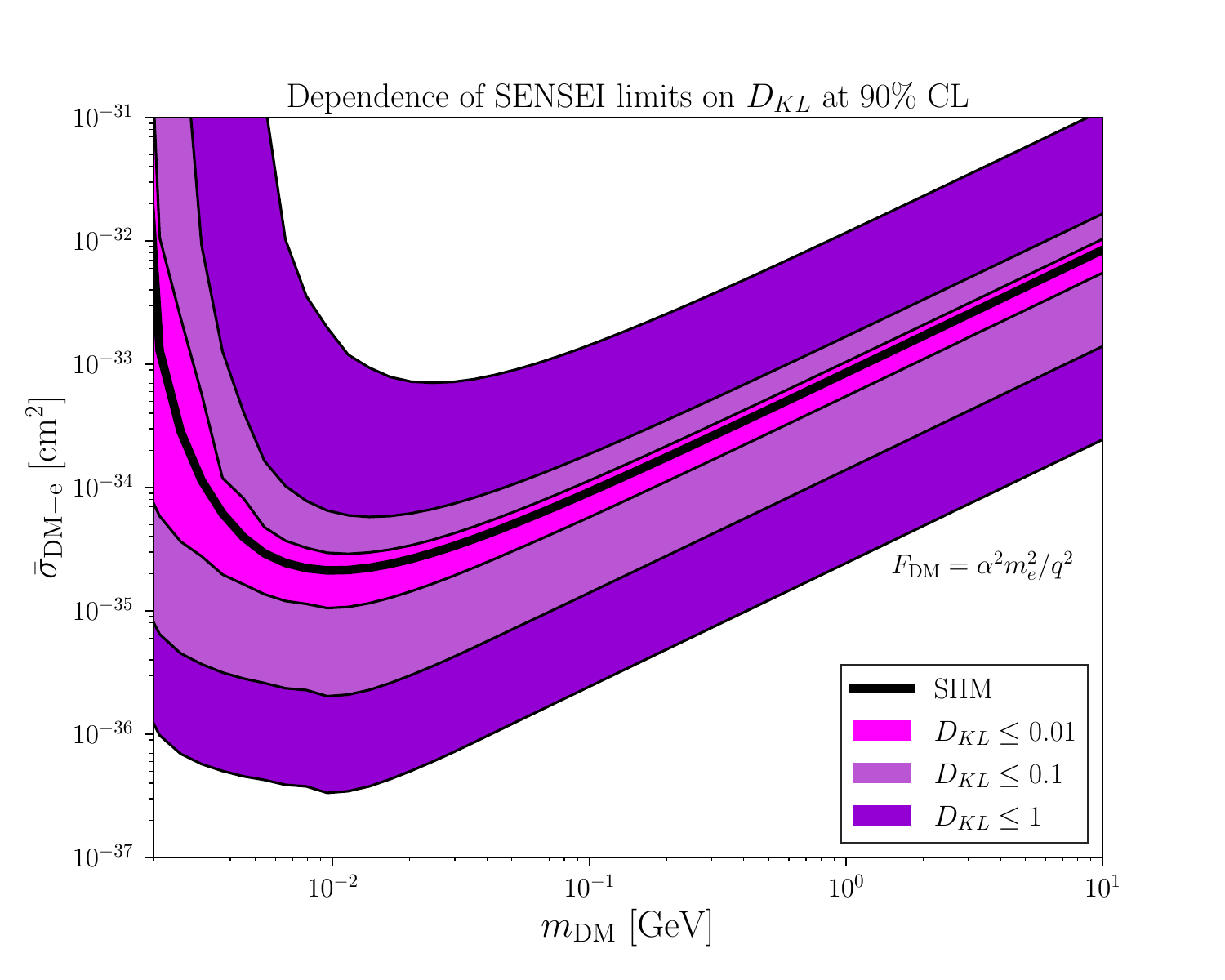}
\includegraphics[width=0.5\textwidth]{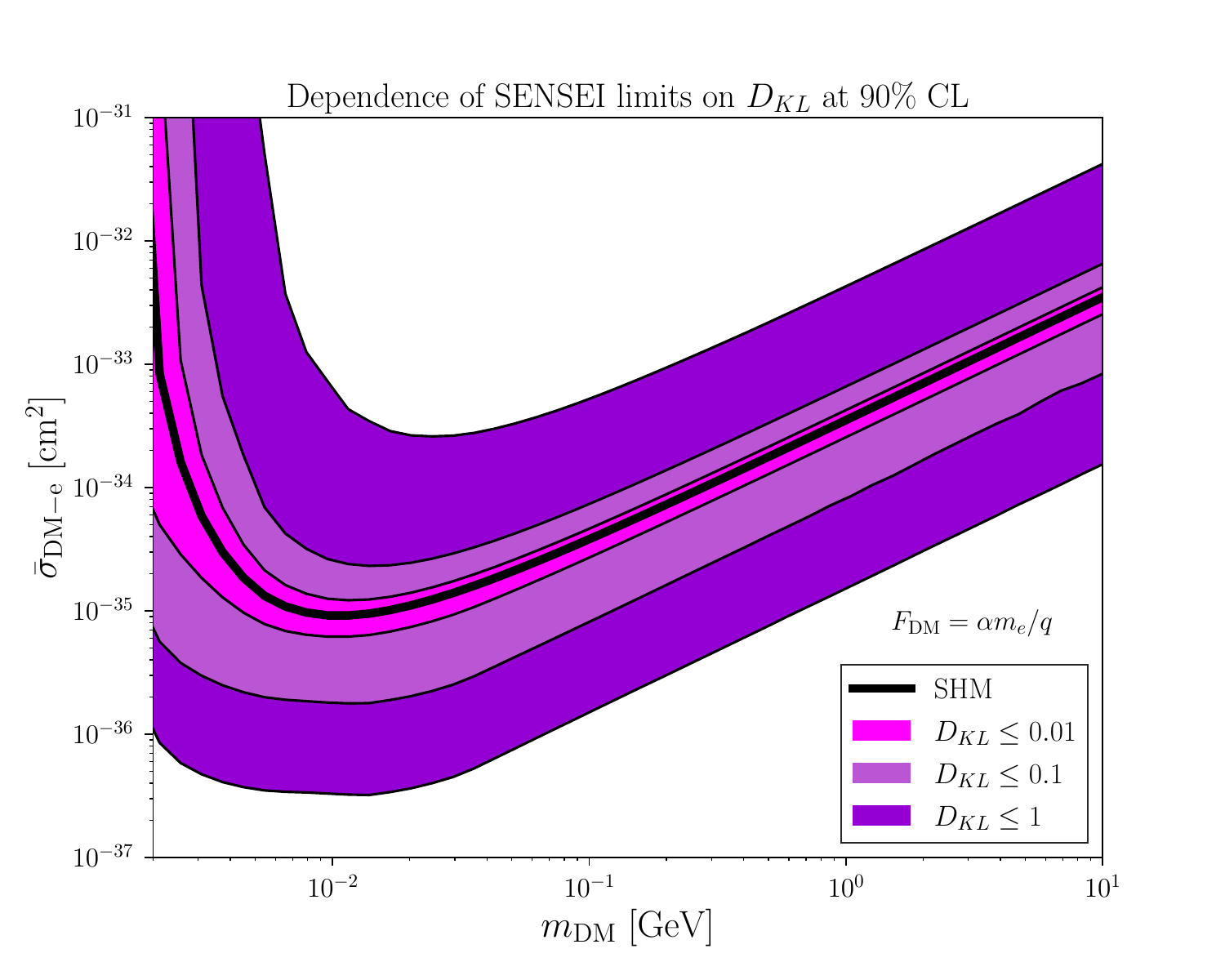}
\begin{center}
\includegraphics[width=0.5\textwidth]{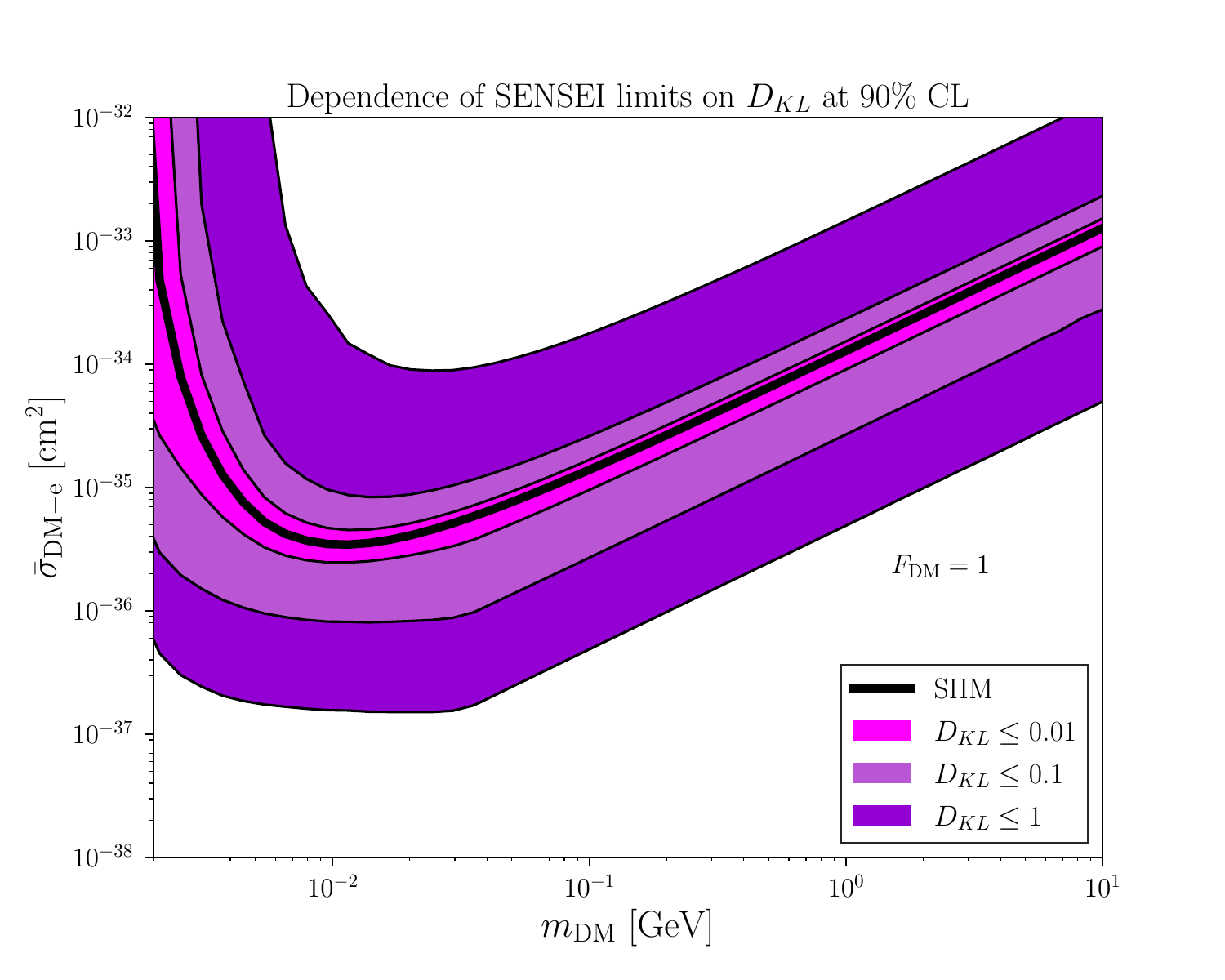}
\end{center}
\caption{Upper limits at 90$\%$ CL from SENSEI, for different maximal deviations (parametrized by the KL divergence) from the Maxwell Boltzmann velocity distribution, and a maximal dark matter velocity of 0.1c.}
\label{fig:limits_Sensei}
\end{figure}

The results from solving this optimization Problem are shown in Figure \ref{fig:limits_Sensei}. We find similar conclusions as for dark matter-nucleon scatterings, although with two main differences. The uncertainties are somewhat larger at low masses, extending over 2 to 3 orders of magnitude for $D_{\rm KL} \leq 1$, and the uncertainties on the velocity distribution manifest more strongly at high dark matter masses than for nuclear recoils, since the electron momentums in the orbital is not fixed. This aspect was discussed previously in \cite{Herrera:2021puj}. The larger impact at low dark matter masses w.r.t nuclear recoils searches in section 5 is not only caused by the electron momentum in the orbitals not being fixed, but also by the restriction on the maximum dark matter velocity done in this Problem. Here we took a maximum value of 0.1c, which allows the optimized velocity distributions to shift towards larger velocities, which have a stronger impact when the dark matter mass is close to the kinematical threshold of the experiment. A smaller choice of $v_{\rm max}$ would reduce the impact of astrophysical uncertainties at low masses assessed with our convex optimization technique.

\begin{figure}[H]
\begin{center}
\includegraphics[width=0.49\textwidth]{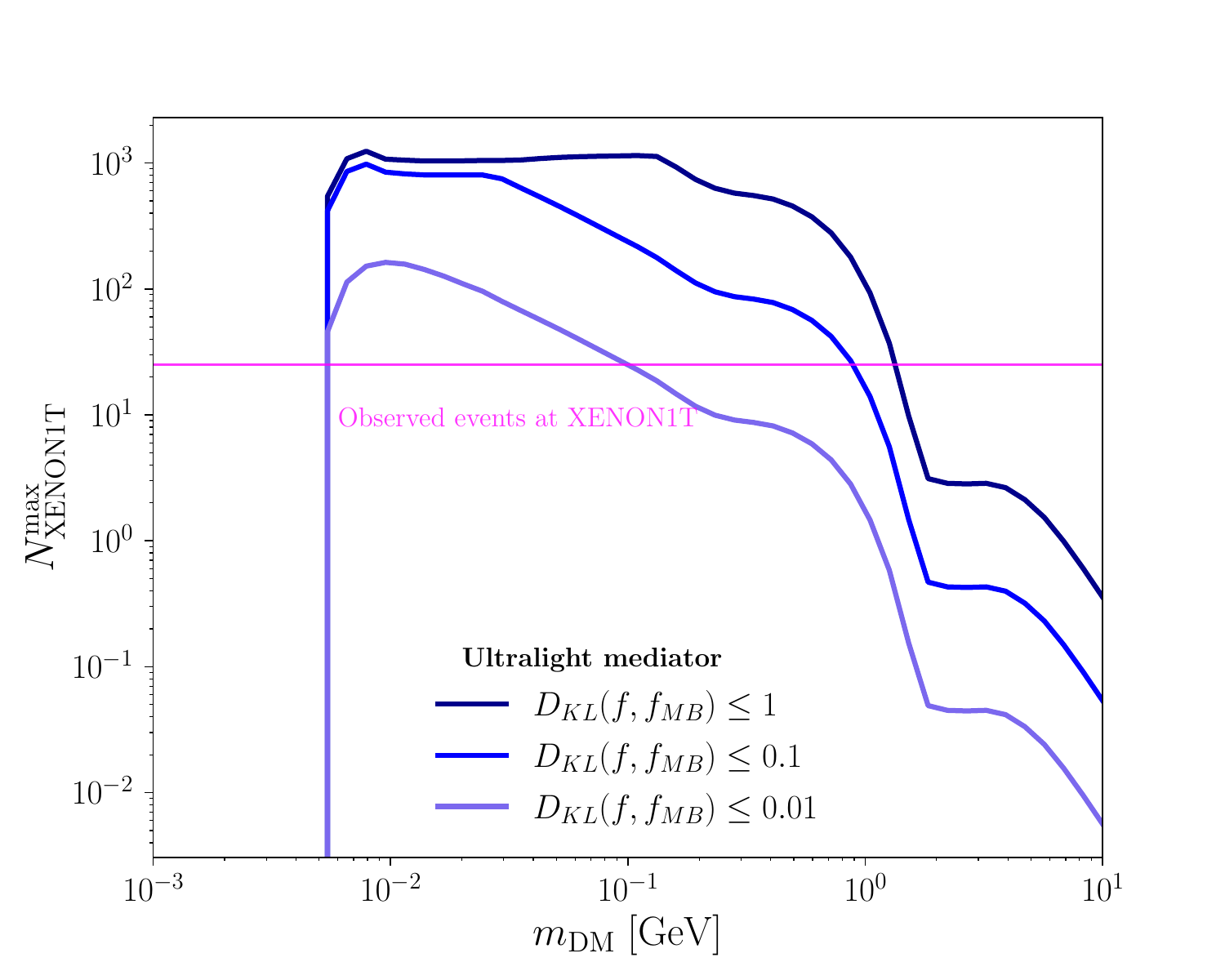}
\includegraphics[width=0.49\textwidth]{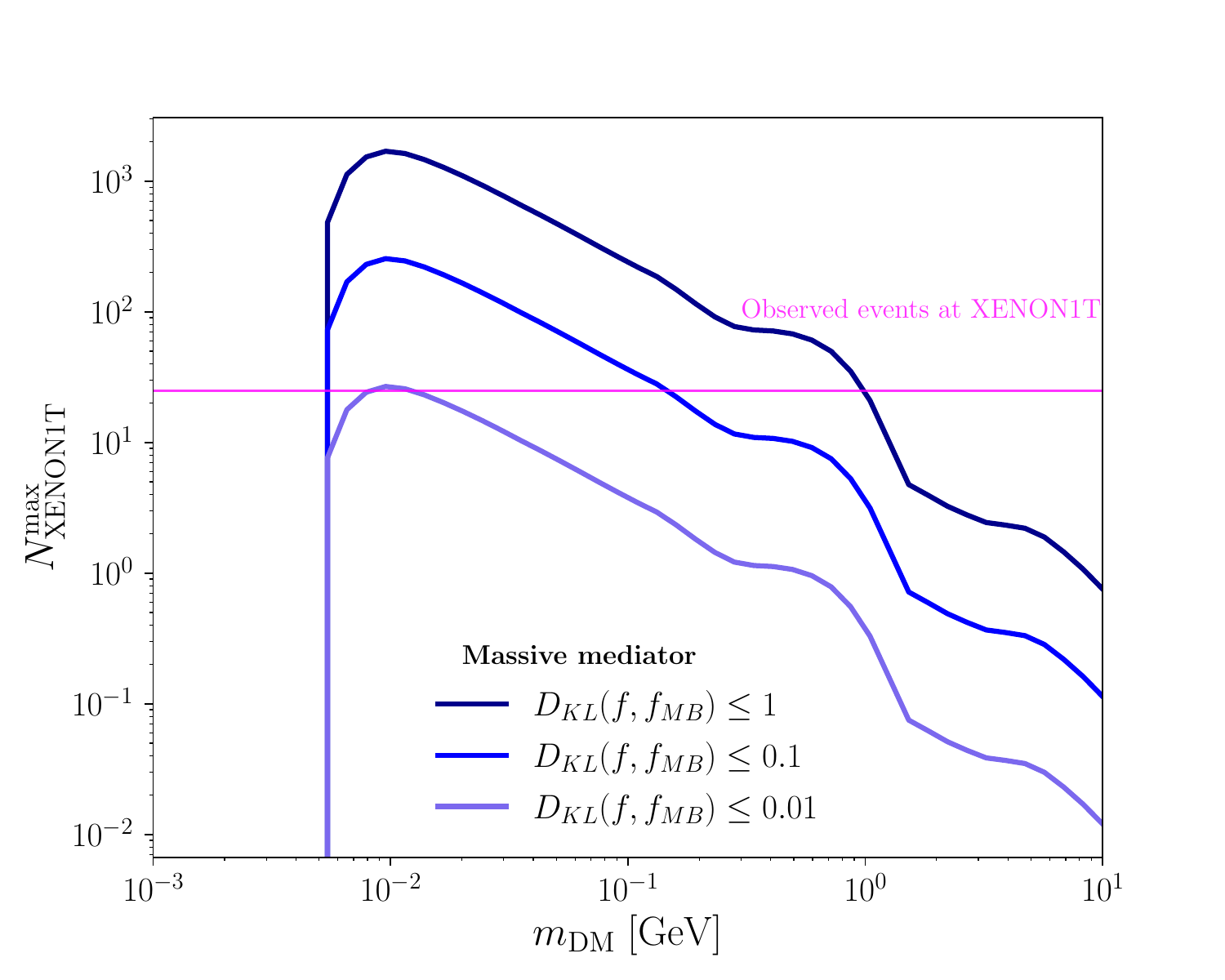}
\includegraphics[width=0.49\textwidth]{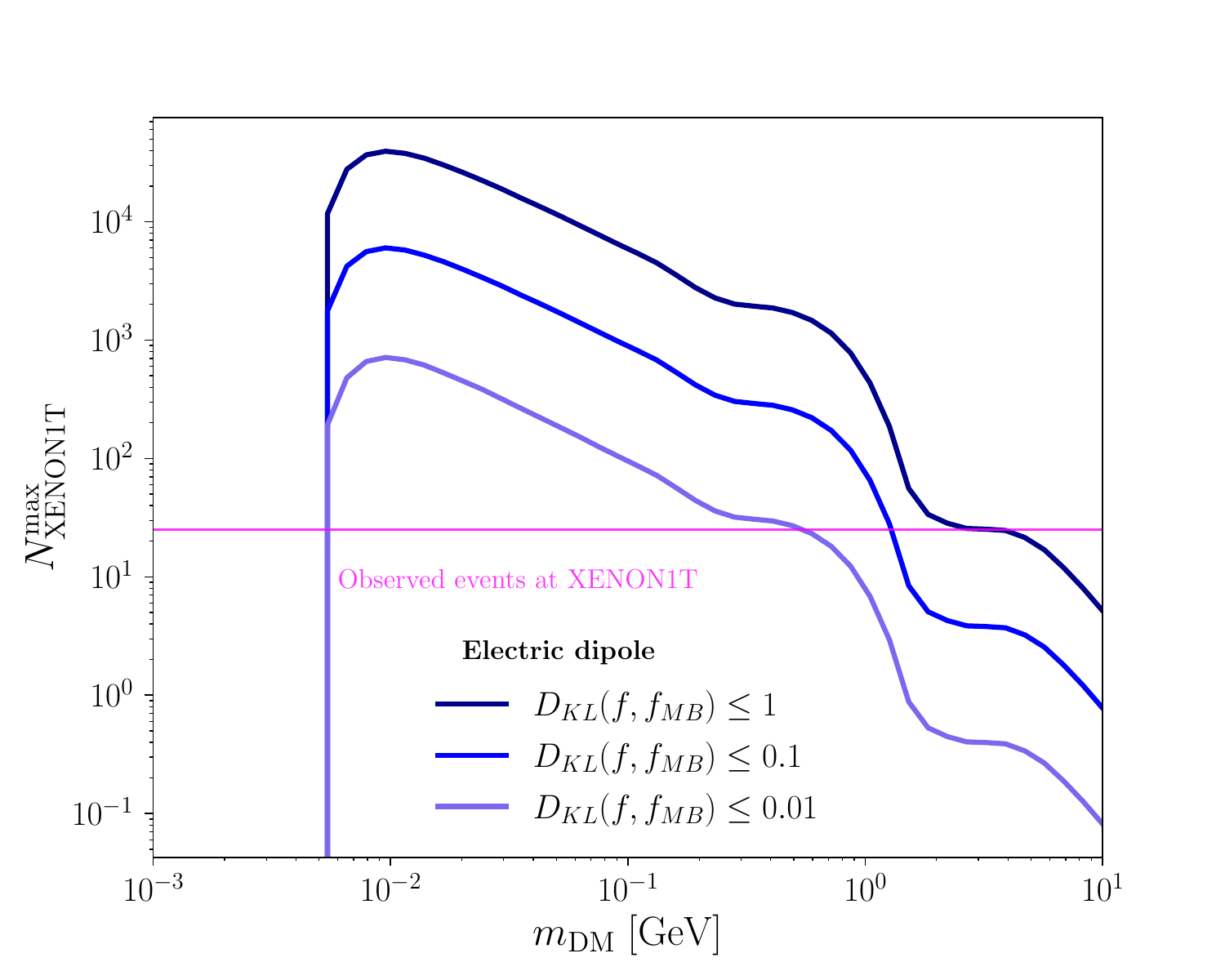}
\end{center}
\caption{Hypothetical optimized number of events in XENON1T, subject to the null results of SENSEI experiment with 10 times more exposure than currently, and from varying the DM velocity distribution w.r.t to the Maxwell-Boltzmann in a model independent manner. For comparison, we show in magenta a hypothetical number of signal events observed in XENON1T, equivalent to the number reported (and later ruled out) in \cite{XENON:2020rca, XENON:2022ltv}.}
\label{fig:XENON_vs_SENSEI}
\end{figure}
We also entertain the possibility to use these methods to confront a future signal in a given dark matter-electron scattering experiment with a null signal from a given experiment. We solve in this case the same optimization problem as described previously, but introducing an additional constraint ensuring that the optimized velocity distribution still gives a null result in one of the experiments.

\clearpage

Concretely,

\begin{center}
\text{\textbf{Maximize:} }
\begin{equation}\label{eq:optproblemdiscrete2}
    \sum_{i} a_{\vec{v}_{i}}R_{ion, \rm XENON1T}^{\vec{v}_{i}}
\end{equation}
\textbf{subject to:}
\begin{equation*}
D_{\rm KL}(f_{\rm MB},f) \leq \Delta
\end{equation*}
\begin{equation*}
\sum_{i=1} a_{\vec{v}_{i}}=1
\end{equation*}
\begin{equation*}            
a_{\vec{v}_{i}} \geq 0
\end{equation*}
\begin{equation*}
    \sum_{i} a_{\vec{v}_{i}}R_{ion, \rm SENSEI}^{\vec{v}_{i}} \leq N_{\rm SENSEI}^{ \rm 90 \% C.L}
\end{equation*}
\end{center}
where we take an upper limit on the number of signal events at SENSEI at 90\% C.L of $N_{\rm SENSEI}^{ \rm 90 \% C.L}=4.957$ events per gram-day of
exposure \cite{Herrera:2023fpq, Barak_2020}.

In Figure \ref{fig:XENON_vs_SENSEI} we show the maximum number of events that can be obtained in XENON1T, assuming an energy threshold from its S2-only analysis data \cite{XENON:2019gfn}, which is 0.18 keV, and assuming the null results from SENSEI with a projected exposure 10 times larger than the current one, which is 48 days $\times$ grams. We do this exercise for different values of the dark matter mass, interaction model and maximum deviation from the Maxwell-Boltzmann. In this way we show that a future signal might be confronted with a null result independently from the assumptions on the halo model, just fixing the maximal deviation from the SHM in a sounded statistical manner.

As can be seen in the plot, the complementarity of XENON1T and SENSEI is clearly more apparent at high dark matter masses, and the ultralight mediator or dipole model shows more complementarity between experiments, since SENSEI threshold to low energy depositions enhances its sensitivity to these models. On the contrary, for a massive mediator, XENON1T sensivity supersedes SENSEI by orders of magnitude, thus, it is more likely than a signal in XENON1T will be observed in SENSEI as well. Indeed, as can be seen in the plot, for values of the KL-divergence below 0.01, a signal would be incompatible with a null result, even with current SENSEI exposures. We also point out that in our figures, the lowest dark matter mass probed is set by the maximal dark matter velocity considered in the optimization problem (0.1 c).

In this discussion, we have not addressed the potential overlap of a direct dark matter-electron signal with an ionization signal from the Migdal effect due to light dark matter-nucleus scatterings \cite{Ibe:2017yqa, Dolan:2017xbu, Essig:2019xkx}. The Migdal effect signal is expected to be subdominant at the XENON1T, but may become larger to the dark matter-electron scattering signal at the energies probed by SENSEI. Moreover, in the future the sensitivity of these experiments may reach the ionization signal induced by neutrino-nucleus or neutrino-electron scattering \cite{Ibe:2017yqa, Bell:2019egg, Herrera:2023xun, Essig:2011nj, Carew:2023qrj}, and the "number of events" counting analysis done here would be incomplete. A more refined analysis accounting for spectral features would become needed. Our convex-opimitizaion technique could be adapted to more sophisticated likelihood analysis.

\section{Conclusions}
	
We have discussed a method to bracket the uncertainty in the theoretical prediction of an observable functional $\mathcal{F}[f]$, consisting in in using methods of convex optimization to sample over all functions $f$ lying within a given distance measure from a central function. In this manner, we account for the uncertainties in the functional $\mathcal{F}[f]$, without assuming any parametric form of the function $f$, as it is typically done in the high-energy physics literature. Our discussion is focused on constraints consisting on distance measures $D(f,f_{0})$ between a reference function $f_{0}$ and the optimized function $f$, given by information divergences.

We applied this method to quantify the impact of the uncertainties in the dark matter velocity distribution in direct detection experiments with information divergences. By using this technique, we have derived upper limits on the spin-independent (spin-dependent) dark matter scattering cross section using CRESST-III and LUX-ZEPLIN (PICO-60) data, accounting for uncertainties in the velocity distribution with different distance measures. We find $\mathcal{O}(1-10)$ uncertainties in direct detection experiments upper limits, for values of the distance measures compatible with dark matter velocity distributions obtained from simulations and observations. We have also derived limits on the dark matter electron-scattering cross section from SENSEI with this technique, sampling over a velocity range extending beyond the escape velocity of the Milky Way. We find uncertainties that are larger than for nuclear recoils, reaching values as high as 3 orders of magnitude, and extending towards larger masses. Finally, we have also explored the potential complementarity of XENON1T and SENSEI in probing a signal vs a null result in the future for different dark matter-electron interaction models. We find a strong complementarity at high masses, while at low masses the experiments are expected to become more degenerate to the dark matter velocity distribution, especially if the mediator of the interaction is ultralight or massless.

Furthermore, we have found that the KL-divergence is a statistically more robust way to account for dark matter astrophysical uncertainties than the relative differences used in \cite{Ibarra:2018yxq}, since it is less constrained by initial assumptions on the shape of the distribution, and the feasible set of velocity distributions that respect this constraint is more varied than for other distance measures. The presence of the logarithm in the KL-divergence causes the optimized velocity distributions to have smoother shapes, that could match physically motivated velocity distributions like those obtained from simulations and observations. However, we emphasize that this methodology does not reconstruct the dark matter velocity distribution from the experimental data, but rather are mathematically optimized distributions that allow for an interpolated model-independent analysis.
	
The parametrization of the uncertainties of a function with information divergences allows to account for uncertainties in physical observables in a novel manner. In some scenarios, a parametric form for the physical function $f$ might be well motivated, and accounting for its uncertainties by varying the value of its parameters could be already be a robust approach. On the other hand, in many other scenarios the true parametric form of the function $f$ could strongly differ from the theoretical prior, and the interpretation of experimental results on the observable $\mathcal{F}[f]$ could be inadequate, given the wrong parametric estimate of $f$. We hope that the techniques present in this work could help a more robust interpretation of experimental results in high-energy and astroparticle physics, particularly in dark matter direct detection searches.

\subsection*{Acknowledgments}
 We are grateful to the Referees for useful comments and suggestions for improvement. We thank Alejandro Ibarra for supervision of this project during the Master and PhD theses of GH and AR, and for providing insightful feedback. GH is grateful to the members of the CRESST collaboration for useful discussions and feedback on direct detection experiments during his Master Thesis, and to Felix Kahlhöfer for help on learning and using the software \texttt{DDCalc}. This work is supported by the Collaborative Research Center SFB1258, and by the Deutsche Forschungsgemeinschaft (DFG, German Research Foundation) under Germany's Excellence Strategy - EXC-2094 - 390783311. The work of GH is also supported by the U.S. Department of Energy Office of Science under award number DE-SC0020262

\appendix
\section{Exponential cone problems}\label{sec:A2}
	Many optimization problems can contain exponentials and logarithms. These can sometimes be modelled with the exponential cone, which is a convex subset of $\mathbb{R}^{3}$ defined as \cite{mosek}
\begin{equation}
K_{\exp }=\left\{\left(x_{1}, x_{2}, x_{3}\right): x_{1} \geq x_{2} e^{x_{3} / x_{2}}, x_{2}>0\right\} \cup \\
\left\{\left(x_{1}, 0, x_{3}\right): x_{1} \geq 0, x_{3} \leq 0\right\}
\end{equation}
Thus the exponential cone is the closure in $\mathbb{R}^{3}$ of the set of points which satisfy
\begin{equation}\label{eq:recastlog}
x_{1} \geq x_{2} e^{x_{3} / x_{2}}, x_{1}, x_{2}>0
\end{equation}
When working with logarithms, as in the case of the KL-divergence, Eq. \ref{eq:recastlog} can be rewritten as
\begin{equation}
x_{3} \leq x_{2} \log \left(x_{1} / x_{2}\right), x_{1}, x_{2}>0
\end{equation}
Alternatively, it can be written as
\begin{equation}
x_{1} / x_{2} \geq e^{x_{3} / x_{2}}, x_{1}, x_{2}>0
\end{equation}
This shows that $K_{exp}$ is a cone, i.e. $\alpha x \in K_{exp}$ for $x \in K_{exp}$ and $\alpha \geq 0$. The convexity of $K_{exp}$ follows from the fact that the Hessian of $f(x,y)=y\exp(x/y)$,
\[
D^{2}(f)=e^{x / y}\left[\begin{array}{cc}
y^{-1} & -x y^{-2} \\
-x y^{-2} & x^{2} y^{-3}
\end{array}\right]
\]
is positive semidefinite for $y>0$.

 
\printbibliography
\end{document}